\documentclass[a4paper,fleqn,usenatbib,useAMS]{mnras}
\usepackage{graphicx}	
\usepackage{amsmath}	
\usepackage{amssymb}	
\usepackage{multirow,multicol}     
\usepackage{bm}		
\usepackage{pdflscape}	
\usepackage[normalem]{ulem}
\newcommand{\degrees}{^{\circ}}
\usepackage[T1]{fontenc}
\usepackage{ae,aecompl}
\usepackage{newtxtext,newtxmath}

\title[A New Catalogue of Galactic Novae]{A New Catalogue of Galactic Novae: Investigation of the MMRD relation \& Spatial Distribution}

\author[\"Ozd\"onmez, A. et al.]{Aykut \"Ozd\"onmez$^1$
\thanks{E-mail: aykut.ozdonmez@ogr.iu.edu.tr}, 
Erg\"un Ege$^{1,2}$, 
Tolga G\"uver$^{2,3}$, 
and Tansel Ak$^3$ \\ 
$^1$Istanbul University, Graduate School of Science and Engineering, 34116, Beyaz\i t, Istanbul, Turkey\\
$^2$ Istanbul University Observatory Research and Application Center, Beyaz\i t, 34119 Istanbul, Turkey\\
$^3$ Istanbul  University,  Faculty of Science,  Department  of
Astronomy and Space Sciences, Beyaz\i t, 34119, Istanbul, Turkey\\
}

\date{Accepted 2018 February 13. Received 2018 February 13; in original form 2017 November 06.} \pubyear{2018}

\begin{document} \label{firstpage}
\pagerange{\pageref{firstpage}--\pageref{lastpage}} \maketitle

\begin{abstract} 
In this study, a new Galactic novae catalogue is introduced collecting important parameters of these sources such as their light curve parameters, classifications, full width half maximum (FWHM) of H$_\alpha$ line, distances and interstellar reddening estimates. The catalogue is also published on a website with a search option via a SQL query and an online tool to re-calculate the distance/reddening of a nova from the derived reddening-distance relations. Using the novae in the catalogue, the existence of a maximum magnitude-rate of decline (MMRD) relation in the Galaxy is investigated. Although an MMRD relation was obtained, a significant scattering in the resulting MMRD distribution still exists. We suggest that  the MMRD relation likely depends on other parameters in addition to the decline time, as FWHM H$_\alpha$, the light curve shapes. Using two different samples depending on the distances in the catalogue and from the derived MMRD relation, the spatial distributions of Galactic novae as a function of their spectral and speed classes were studied. The investigation on the Galactic model parameters implies that best estimates for the local outburst density are 3.6 and 4.2 $\times 10^{-10}$ pc$^{-3}$ yr$^{-1}$
with a scale height of 148 and 175 pc, while the space density changes in the range of $0.4 - 16 \ \times 10^{-6}$ pc$^{-3}$. 
The local outburst density and scale height obtained in this study infer that the disk nova rate in the Galaxy is in the range of $\sim20$ to $\sim100$ yr$^{-1}$ with an average estimate $67^{+21}_{-17}$ yr$^{-1}$.
\end{abstract}


\section{INTRODUCTION}
Novae are defined as systems with the largest outburst amplitude among
other sub-types of cataclysmic variables, which is a short-period
interacting binary system containing a white dwarf (WD) and a donor star
\citep{2003cvs..book.....W}. A nova outburst occurs on the surface of a white
dwarf where material accumulates from a donor star until the pressure
and temperature are high enough to trigger a thermonuclear runaway that ejects
the accreted envelope \citep{2008clno.book.....B, 2014ASPC..490.....W}. 
Systems with much smaller recurrence times, called recurrent novae (RNe), show at least two outbursts in the
observational history, while systems classified as classical novae
(CN) have only one outburst discovered. The relatively massive WDs and
higher mass transfer rates in RNe probably shorten the recurrence
time \citep{1985ApJ...291..136S, 2005ApJ...623..398Y}. Traditionally, the classification of novae is based on outburst properties. They are classified by their speed class \citep{1957gano.book.....G}, determined via their decline times, as fast or slow novae, by their spectral class \citep{1992AJ....104..725W}, determined via the form of the early outburst spectra, as Fe II, He/N or hybrid novae. \citet{2010AJ....140...34S}
classified Galactic novae based on the light curve shapes into
smooth (S), plateau (P), dust dip (D), cusp (C), oscillations (O), flat
topped (F), and jitter (J). Besides, the classifications based on the outburst properties, 
these systems can also be classified via the evolutionary state of
the donor star \citep{2012ApJ...746...61D} into main sequence, sub-giant or
red giant star novae; or via the population of the system in the host galaxy
\citep{1998ApJ...506..818D} into disk or bulge novae.

Novae are important sources to study for a number of reasons. 
For example, novae, particularly, RNe that host a high-mass WD (likely close to the
Chandrasekhar limit) with sub-giant or red giant donor star, are
considered as potential progenitors of type Ia supernovae,
which are used to measure the accelerating expansion of the Universe
\citep{2014ARA&A..52..107M}. They are also important sources for the enrichment 
of the interstellar medium in carbon, nitrogen, oxygen and lithium 
\citep{2009ApJ...692.1532S, 2015Natur.518..381T}. They
are X-ray and high energy gamma-ray sources \citep{2014Natur.514..339C}. 
Moreover, nova eruptions also allow studies of the
common-envelope process and of the shock physics \citep{2014Natur.514..339C}.

With the advances in recorded nova outbursts and obtained data over the last
decades, the number of novae with known fundamental parameters
increased, and their outburst mechanism, nature of light curves and
spectra became more understandable. For example,
\citet{2010AJ....140...34S} represented well-determined light curves 
including peak magnitudes and dates, decline times, the quiescent
magnitudes, and their light curve classes. To quantify RN candidates which have been classified as CN,  
\citet{2014ApJ...788..164P} analyzed the Galactic nova
sample based on the information on the light curves and spectra, and
recognized RN candidates since they have smaller outburst amplitudes, larger
orbital periods, infrared colors indicating the existence of giant donor stars, high
expansion velocities from full width half maximum (FWHM) of H$_\alpha$ line at 6563 {\AA}, high excitation lines such as
\ion{Fe}{x} or \ion{He}{II} near outburst peak, a light curve shape class of P-type, and WD
masses greater than 1.2M$_{\odot}$. From the new observations, the number of discovered bulge 
Galactic novae were also increased \citep{2015ApJS..219...26M}, and the nova rate was calculated both for disk and bulge systems \citep{2017ApJ...834..196S,2015ApJS..219...26M}.  

Recently, we determined the distances of a large sample of novae using a systematic approach \citep{2016MNRAS.461.1177O}, which depends on the reddening-distance
relation derived by utilizing the unique location of red clump stars on
colour-magnitude diagrams from 2MASS \citep{2006AJ....131.1163S}, UKIDSS \citep{2008MNRAS.391..136L} and VISTA VVV \citep{2012A&A...537A.107S} near-infrared surveys. 
The results are consistent with 
the distances inferred from parallax method within the stated errors, and the method has the advantage of being applicable to a large number of novae. Hence, the existence of a maximum magnitude - rate of decline  (MMRD) 
relation for the Galaxy, which has long been questioned and has been used to calculate
absolute magnitudes of novae, can be investigated with these new
distances and light curve parameters.

There have been numerous characterizations of the MMRD relation for
Galactic \citep[e.g.][]{1985ApJ...292...90C, 2000AJ....120.2007D} and extragalactic nova
populations \citep[e.g.][]{1990ApJ...360...63C,1995ApJ...452..704D,2006MNRAS.369..257D}.
A recent study by \citet{2011ApJ...735...94K} and \citet{2017ApJ...839..109S} has revealed a number of apparently
faint, yet relatively rapidly fading novae in M31 and M87, which caused the
authors to question whether an MMRD relation is justified at all.
Theoretically, the WD mass determines the main trend of the MMRD relation, and 
the ignition (initial envelope) mass, in other words, the mass-accretion
rate to the WD causes the scatter in the MMRD relation 
\citep{2005ApJ...623..398Y, 2010ApJ...709..680H}. If the mass-accretion rate 
to the WD is relatively
larger, the ignition mass is smaller, hence the peak brightness is
fainter. However, the heterogeneity in nova light curves suggests 
that a single parameter may not characterize the decline well. Thus, other
properties such as spectroscopic behavior, light curve shape which
characterize novae in different ways may resolve the
scatter in the MMRD relation.

In this study, we compile a new Galactic nova catalogue, which provide information on the nova population as described in Section \ref{SecCat}. Using the data in the catalogue, we investigate absolute magnitude-decline time distributions to obtain the MMRD relation of the Galactic novae in Section \ref{SecMMRD}. The statistical analyses of the Galactic novae on the decline times, absolute magnitudes at maximum outburst, and spatial distributions are given in \ref{SecSpatial}. We also investigate Galactic model parameters as space density and scale height in the same section. Using the obtained Galactic model parameters, we also calculated the nova rate of disk novae. Finally, Section \ref{SecDisCon} represents a review of the results with a discussion of the possible uncertainties.

\section{The Catalogue for the Galactic Novae}
\label{SecCat}
We compile a new catalogue of Galactic novae for which the distance information is acquirable. The catalogue contains 291 novae and almost all their important parameters such as their light curve parameters (amplitude A, maximum
magnitude at outburst $V_{max}$, decline times $t_2,t_3$), shape of the light curve, spectral, recurrence classifications, FWHM H$_\alpha$ and orbital period $P_{orb}$.
The distances and/or interstellar reddening estimates are crucial parameters to obtain the luminosity function, spatial distributions or absolute physical parameters of these binary systems. In the catalogue,
we included all the distance information of more than 150 novae in  the literature obtained
from the expansion/trigonometric parallaxes or from reddening-distance relations
\citep[RDRs; ][]{2016MNRAS.461.1177O}. The distances for the remaining novae could be obtained from
MMRD relations. Besides, all reddening estimations for these nova systems were collected 
from the literature together with their measurement methods. 
 
The distance measurements of the novae in the catalogue depends on two approaches; parallax and RDR methods. The most reliable method is the trigonometric parallax, but the number of novae with trigonometric parallax is less than 10 in the entire observational history even with GAIA DR1 \citep{2016A&A...595A...1G}. 
Distances can also be calculated by following the expansion parallax method which depends on the angular expansion of the resolved nova shell. However, it is not free from systematic biases due to the complex structure of shells, non-uniform shell expansion, inconstant expansion velocities etc. In the catalogue, $\sim 30$ novae have expansion parallaxes. For some novae, e.g. LW Ser, QV Vul, LV Vul, the distances from expansion parallaxes, determined at different phases of the outburst and/or from the profiles of various spectral lines, show significant scatter.
In the previous study \citep{2016MNRAS.461.1177O}, we utilized a new distance
calculation method for these systems via RDRs that helped us to
determine the distances for nearly all known Galactic novae at low Galactic latitudes.
In this RDR method, the uncertainties of the distance measurements arise from
reddening estimates of novae as well as the shallow increase in the RDRs in some directions of the Galaxy. The comparison of the distances calculated following this method with that from the parallax method in \citet{2016MNRAS.461.1177O} showed that two methods are in good agreement within the stated errors, except for LV Vul, LW Ser, and QV Vul.

There are a number of studies that compile data 
from AAVSO\footnote{https://www.aavso.org/} (American Association of Variable Star Observers) and/or from literature to present the light curve
properties of Galactic novae, e.g. \citet{1987MNRAS.227...23W, 1997ApJ...487..226S,
2008AstL...34..241B}, but a large number of qualified light curves with
extensive time coverage (often until quiescence) that go deep and have many
observations each day are only presented by \citet{2010AJ....140...34S}. As
mentioned in \citet{2010AJ....140...34S}, the compilations, which include
observations before the AAVSO data started, 
missed the peak of almost half of
the novae, and correspondingly the decline times ($t_2,t_3$) are not well determined.
Besides, adopting a maximum brightness ($V_{max}$) and decline times during nova outburst without comprehensive knowledge of the light curve shape of a nova may be erroneous.  
Hence, the light
curve shapes of novae in our sample were additionally checked 
using the AAVSO archive and/or related studies of the nova in question. 
If the novae have enough observations to clearly represent declining in the light curve, we assumed that the light curve parameters of the novae are adequately reliable.  
Recently, \citet{2014ApJ...788..164P} analyzed the Galactic nova
sample from the information on the light curves and spectra to 
quantify RN candidates. FWHM H$_\alpha$ around the
outburst maximum is one of the important parameters along with the amplitude of the outburst, decline times, light curve shapes and excitation lines that correspond to the recurrence of a novae \citep{2014ApJ...788..164P}. 
The expansion velocity of the ejected material in a nova
outburst is measured using widths of emission lines, generally using the H$_\alpha$ line.
It may also help to clarify the scatter in the MMRD relation as discussed in section \ref{SecMMRD}.

Our catalogue is published on the website\footnote{http://highenergyastro.istanbul.edu.tr/novae\_cat}.
For 189 Galactic novae, the RDRs were derived from the method described in \citet{2016MNRAS.461.1177O}, and these relations are shown at individual webpages. 
It also contains an algorithm to reproduce calculations of either the distance or the reddening for a given value, from derived RDRs. The website contains a main table in which the
adopted parameters are given and an SQL based query can be performed to search novae systems. All 
references and detailed descriptions for the parameters are given on the website as well.
The catalogue will be updated frequently with most recent determined parameters and new RDRs for the novae, which were not investigated.

We listed the novae in the catalogue in two tables. 
The novae with distances calculated from parallaxes or
RDRs are presented in Table \ref{table1}. The novae used 
to derive the MMRD relation are given in Table \ref{table1}a. Unreliable or 
questionable parameters for the novae, which are not used in the derivation of the MMRD 
relation, are listed in Table \ref{table1}b.
The remaining novae that do not have any distance measurement are listed in Table 
\ref{table2}. The light curve parameters of the novae in Table \ref{table2}a 
are well-determined to allow the accurate measurements from the MMRD relation.

\section{The MMRD Relation of Galactic Novae}
\label{SecMMRD}
In this section, we aim to analyze the existence of the MMRD relation for the
Galactic novae that has long been questioned. In order to analyze MMRD relations,
we used only well-determined light curve parameters and distances as given 
in Table \ref{table1}a.
Some of the novae (e.g. CK~Vul, CT~Ser, V2674~Oph) 
do not have enough observations to obtain
a reliable maximum magnitude and/or decline rates from outburst light-curves, and are not included in the analyses. Such novae are given in Table \ref{table1}b. We also give information on the reason of why a nova in Table \ref{table1}b is not
used for the investigation of the MMRD relation. For example, the distances of the novae BY~Cir, V1301~Aql and V1493~Aql obtained from RDRs vary largely owing to different extinction
measurement methods. Even a small variation in the reddening estimate changes the distance 
calculation drastically, for instance in case of V394~CrA, Q~Cyg, V373~Sct, 
V4160~Sgr, V5116 Sgr. For some other novae (e.g. V394~CrA, V368~Sct), more reliable reddening
measurements are required to calculate accurate distances using this method. 
Since the distance calculation is strongly correlated with the reddening estimate, we also
mark the novae where the distance calculation depends on only one reddening estimate.  

\begin{figure*}
\includegraphics[width=0.9\textwidth]{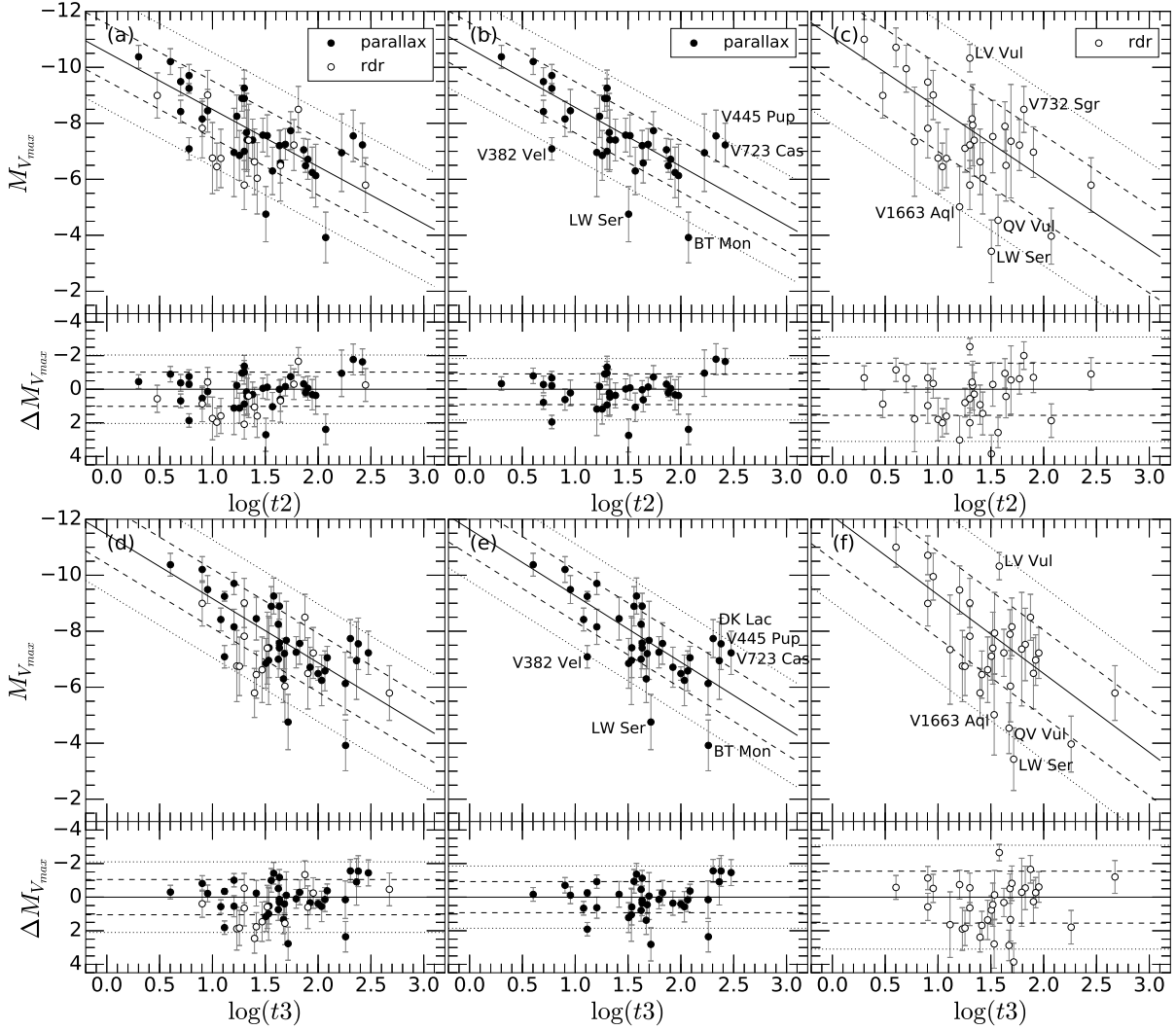} 
\caption{$M_{V,max} - t_{2, 3}\ $ distributions of Galactic novae: (a,b,c) and (d,e,f) for $t_2$ and $t_3$, respectively. (a, d) for the novae with distances obtained from either parallax or RDR , (b, e) only with parallax measurements and (c, f ) only with distance measurements from RDR. In the upper panels, the black lines represent the MMRD relation, the dashed and dotted lines show an uncertainty of $\pm1\sigma$ and $\pm2\sigma$ of the fit, respectively. The residuals are shown in the lower panels. The names of novae with relatively larger scattering are given.}
\label{fig:MMRD} \end{figure*}

\begin{figure*}
\includegraphics[width=0.9\textwidth]{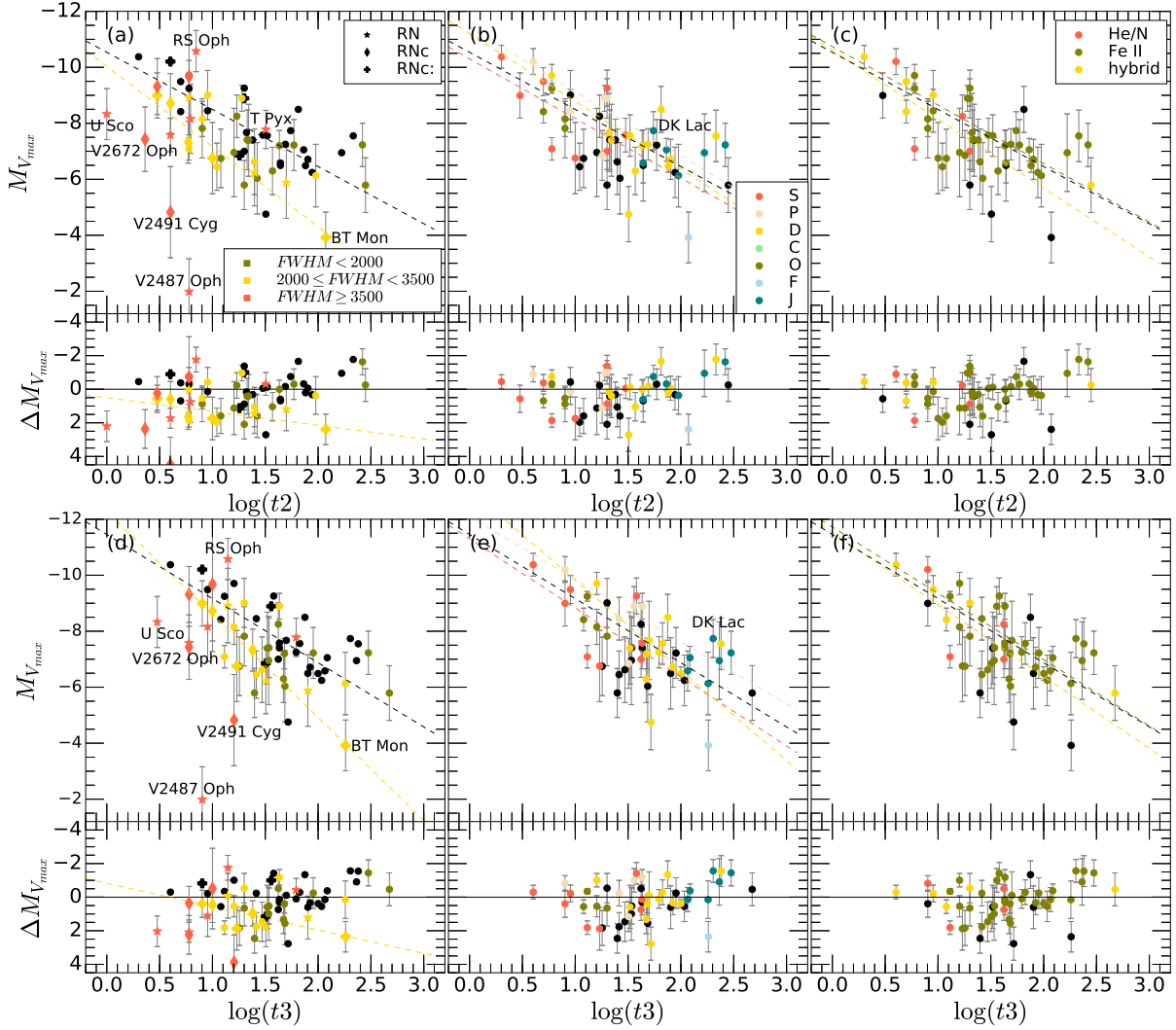} 
\caption{$M_{V,max} - t_{2, 3}\ $ distributions of Galactic novae: (a,b,c) and (d,e,f) are for $t_2$ and $t_3$, respectively, (a,d) considering FWHM H$_\alpha$ and including RN + RN, (b, e) considering the classification of light curve shapes, and (c, f) considering the spectral classification. Black lines represent the adopted MMRD relation. We denoted the trends in the distributions by considering their FWHM H$_\alpha$ velocities, light curve shapes and spectral classification as colored dashed lines, as defined in legends. The residuals are shown in the lower panel. Recurrent novae and  relatively scattered novae are labeled. DK Lac changes its position on these MMRD distributions owing to a non-linear decline.}
\label{fig:MMRD_other_spec} \end{figure*}

Using the Galactic nova samples in which the distance measurements depends on two different approaches (parallax and/or RDR),
we plot $M_{V,max} - t_{2, 3}$ distributions
(see, Fig. \ref{fig:MMRD}) to search for MMRD relations. For the main sample, we preferred to use the distances from the parallax method switching to that from RDRs. The other sample contains the novae whose distances are calculated using a particular distance measurement method. Hence, we test systematic differences between MMRD relations obtained from different distance measurement methods.
On the other hand, a number of studies that are based on theoretical models, suggest that the classical MMRD relation does not work properly for RNe, as seen in Fig. \ref{fig:MMRD_other_spec}. For these reasons, we analyzed the distributions by separating the RNe and RN candidates from classical novae. 
Note that while obtaining MMRD relations, we excluded only
strong RN candidates given in \citet{2014ApJ...788..164P}, for which many of the indicators strongly point to them being recurrent but only one outburst has been detected.

The uncertainty in the absolute magnitude at outburst maximum was estimated by accounting for the uncertainty of the distances and extinction, but for the calculation of the total uncertainty, the uncertainty in the maximum magnitude should also be taken into account. However, the brightest observed V-band magnitudes are usually adopted for the maximum magnitude during the outburst, and it may lead to an erroneous estimate in case the nova was detected after already passing the true maximum. This observational effect may lead to a larger uncertainty than the typical instrumental errors on maximum magnitudes, and therefore the uncertainty of the maximum magnitudes has usually not been considered in the MMRD studies. 
Similar problems exist in defining the decline times that are usually obtained by binning or interpolating the light curves for a given maximum magnitude. There can be multiple measurements of decline times for a light curve that fluctuates around the given magnitude. Thus, we only considered the novae with well-observed light-curves, and we adopted $\pm10\%$ of the maximum magnitude value as the error to represent systematic uncertainties. Moreover, we investigated $M_{V,max} - t_{2, 3}\ $ distributions by considering the classification of light curve shapes to analyze the dependence on MMRD relation due to the non-linear decline of novae possibly arising from jitters, cusps, and dips in light curve, at the end of this section.

We obtained MMRD relations for $t_2$ and $t_3$ decline times with the use of three different novae samples according to the considered distance measurement methods. 
The MMRD relations appear almost linear; $M_{V,max}= a+ b \times \text{log}t_{2, 3}$. The coefficients of the fit and the result from regression analyses by taking into account uncertainties in only the y-axis are given in Table \ref{MMRD_table}.
The coefficients of the MMRD relations obtained from the three different samples of novae are within the stated errors, but the relations obtained from the parallax+RDR sample for both $t_2$ and $t_3$ times have smaller slopes (coefficient b) and larger y-intercepts (coefficient a) than that from the other samples. Even though the novae in the RDR sample have larger errors and show a larger scattering in absolute magnitude at maximum, the coefficients are similar to those of the other relations. 
Thus, we assume that there is no systematic bias between the derived relations when different distance measurement approaches are used.
The reduced $\chi^2$ is expected to be less than 1 for considering model to be over-fitting data, but the the $\chi^2_{red}$ of the MMRD relations obtained in this study are greater than 2. In addition, the $\pm1\sigma$ is about 1 mag, while greatest deviations from MMRD relations are about $2$ mag. Nevertheless, a linear equation appears to be an adequately (but not fully) fit to the data. We used the MMRD relations from the parallax+RDR sample in the further investigations, since the relations were obtained from a larger number of novae, and have smaller standard errors and  $\chi^2_{red}$.
Our results are similar to the most
popular MMRD relation of \citet{2000AJ....120.2007D}, in which the $1\sigma$ scatter 
was $\pm 0.6$ mag with a deviation as large as 1.6 mag.

\begin{table}
\caption{The results of regression analyses for deriving MMRD relations according to $t_2$ and $t_3$ times. Here, N is number of novae in the sample, a and b are coefficients of the relation, y-intercept and slope, $\sigma$ is standard error of the residuals, $\chi^2_{red}$ is reduced chi-squared.}
\label{MMRD_table}
\begin{tabular}{l|c|l|l}
Sample & N & for $t_2$ & for $t_3$ \\
\hline
\multirow{4}{*}{Parallax + RDR} & \multirow{4}{*}{50} & 
$a=-10.54 \pm 0.30$ & $a=-11.46 \pm 0.42$ \\
 & & $b=2.04 \pm 0.23$ & $b=2.29 \pm 0.27$ \\  
 & & $\sigma=0.9$ mag & $\sigma=1.0$ mag \\
& & $\chi^2_{red}=2.14$ & $\chi^2_{red}=2.34$ \\
\hline
\multirow{4}{*}{Parallax} & \multirow{4}{*}{36} & 
$a=-10.68 \pm 0.32$ & $a=-11.64 \pm 0.46$ \\
& & $b=2.11 \pm 0.25$ & $b=2.37 \pm 0.30$ \\ 
& & $\sigma=0.8$ mag & $\sigma=0.8$ mag \\
& & $\chi^2_{red}=2.30$ & $\chi^2_{red}=2.54$ \\
\hline
\multirow{4}{*}{RDR} & \multirow{4}{*}{32} & $a=-11.08 \pm 0.78$ & $a=-12.11 \pm 1.04$ \\
 &  & $b=2.53 \pm 0.59$ & $b=2.80 \pm 0.67$ \\
 &  & $\sigma=1.5$ mag & $\sigma=1.5$ mag \\
 &  & $\chi^2_{red}=2.91$ & $\chi^2_{red}=2.64$ \\
\hline
\end{tabular}
\end{table}

The previous studies on the MMRD relation show a common problem that  there is
considerable intrinsic scatter in the Galactic MMRD relation and this may indicate
the presence of the hidden second-order parameters as mentioned in
\citet{2000AJ....120.2007D}. Thus, we investigated $M_{V,max} - t_{2, 3}\ $ distributions (Fig. \ref{fig:MMRD_other_spec})
considering other properties such as spectroscopic type, 
light-curve shape, and FWHM H$_\alpha$. The median values of the differences between the observed and the calculated absolute magnitudes, $\bar{\Delta M_V(t_{2,3})}$, are also given in Table \ref{absmag_class} in which the novae are separated into samples. The median values for the $t_2$ and $t_3$ times are increasing so long as FWHM $H_\alpha$ gets higher. For a further investigation, we added the RN and candidate RN (cRN) sample into the MMRD distributions (Fig. \ref{fig:MMRD_other_spec}a,d) by considering the FWHM H$_\alpha$. Note that, we used the distances in \citet{2010ApJS..187..275S}  and 
\citet{2013ApJ...773...55S} for well-known Galactic RNe 
(RS Oph, U Sco, V3890 Sgr, and T Pyx), if it did not depend on the MMRD relation, 
since their distances could not be calculated from expansion parallaxes or
RDRs. It is important to reveal the positions of RNe 
in the MMRD distributions corresponding to FWHM H$_\alpha$, since the novae with high velocities seem to be related with RNe \citep{2014ApJ...788..164P}. The absolute magnitudes of RN + cRN
which have $2000 \ \text{km/s}< \text{FWHM H}_\alpha$ are generally fainter than those
obtained from MMRD relations, and the systems with $2000 \leq \text{FWHM H}_\alpha < 3500$ km/s seem to trace different lines: 
$-10.03(\pm0.45)+2.9(\pm0.3) \ \text{log}t_2$ and $-12.3 (\pm0.55)+3.7(\pm0.4)\ \text{log}t_3$. For $3500 \text{ km/s } \leq \text{FWHM H}_\alpha$, the trends are almost vertical, but we did not find significant relations from regression analysis owing to large scattering.
Note that, we adopted the intrinsic errors of the distances of RNe, but the distances of some RNe differ strongly in the literature.
We also tried to obtain MMRD relations by separating novae according to their light curve shape and spectral classifications,
but they are all consistent with those obtained using all novae when taking into account the error of the coefficients. The following relations were obtained for samples in which number of novae is greater than 5;
\begin{itemize}
\item{\makebox[2.5cm][l]{for S:}$-10.3(\pm 0.9) + 2.1 (\pm 1.1) \log t_2$}
\item[]{\makebox[2.5cm][l]{}$-11.3(\pm 1.3) + 2.5 (\pm 1.1 \log t_3$}
\item{\makebox[2.5cm][l]{for P:}$-11.4(\pm 1.2) + 2.3 (\pm 1.2) \log t_2$}
\item[]{\makebox[2.5cm][l]{}$-12.0(\pm 1.6) + 2.2 (\pm 1.2) \log t_3$}
\item{\makebox[2.5cm][l]{for D:}$-11.3(\pm 0.7) + 2.4 (\pm 0.4) \log t_2$}
\item[]{\makebox[2.5cm][l]{}$-13.1(\pm 1.2) + 3.3 (\pm 0.7) \log t_3$}
\item{\makebox[2.5cm][l]{for Fe II:}$-10.7(\pm 0.3) + 2.1 (\pm 0.2) \log t_2$}
\item[]{\makebox[2.5cm][l]{}$-11.7(\pm 0.5) + 2.4 (\pm 0.3) \log t_3$}
\item{\makebox[2.5cm][l]{for He/N+Hybrid:}$-10.6(\pm 0.8) + 2.4 (\pm 0.9) \log t_2$}
\item[]{\makebox[2.5cm][l]{}$-11.5(\pm 1.0) + 2.3 (\pm 0.8) \log t_3$}
\end{itemize}
The MMRD relations for subsamples are all similar with only small differences in the coefficients. Only the relation for the D-type could be slightly different one. In contrast, the novae with J-type light curves remain at a constant absolute magnitude $M_{V,max}\sim-7\pm0.2$ mag for $40 \text{ days } \lessapprox t_{2}$ and $100 \text{ days } \lessapprox t_{3}$ (Fig. \ref{fig:MMRD_other_spec}b,e). 
We also investigated the median values of the differences between the calculated and the observed absolute magnitudes (Table \ref{fig:hist_MV}).
The novae having plateaus (P) or jitters (J) in their light curves have brighter absolute magnitudes than that calculated from MMRD relations, while the novae with oscillations (O) in their light curve have fainter absolute magnitudes. The other nova subsamples of smooth (S) or dips (D) light curve type are consistent when considering the stated errors. V2491 Cyg and BT Mon are the only systems with cusp (C) and flat-topped (F) light curve shape, respectively. In the following section \ref{sec:dec_times_abs}, we investigated the $t_2$ and $t_3$ decline times for all novae in the catalogue, and derived a relation between logarithmic decline times. Moreover, we found that D and J-type novae may not have linear decline between $t_2$ and $t_3$ times.
In that case there may be a dependence on the decline rates $t_2/t_3$ that changes the position of novae in these two MMRD distributions, especially for novae with D- or J-type. Note that, in the MMRD sample, there is only one nova, DK Lac, which is not in the $\pm2\sigma$ limits of that relation. Finally, the mean values of the differences between absolute magnitudes for the subsamples distinguished by the spectral classification are all within the stated errors, but the RNe systems in the He/N class tend to have fainter absolute magnitudes.
However given the fact that the number of systems where these parameters could be determined is very small to confirm any dependence of the decline of the light curve or spectral classification (especially for He/N+hybrid novae) on MMRD relations.

\begin{table}
\caption{The median values of the differences between the calculated and the observed absolute magnitudes for the subsample, separated according to FWHM $H\alpha$ measurements, light curve shapes (F, P, S, J, C, D), and spectral classes (He/N, Fe II, hybrid). Here, N is the number of novae in the sample, $\bar{\Delta M_V}(t_2)$ and $\bar{\Delta M_V}(t_3)$ are the differences of the absolute magnitudes when using $t_2$ and $t_3$ times, respectively. LF, MF and HF represent the novae with low FWHM $H_\alpha< 2000$ km/s, medium  $2000\leq$ FWHM $H_\alpha< 3500$ km/s and high FWHM $H_\alpha\geq 3500$ km/s, respectively.}
\label{absmag_class}
\small
\begin{tabular}{l@{\hskip4pt}c@{\hskip4pt}c@{\hskip4pt}c@{\hskip4pt}c@{\hskip4pt}c@{\hskip4pt}c@{\hskip4pt}c@{\hskip0pt}}
\hline
 &  \multicolumn{3}{c}{CNe}  &  \multicolumn{3}{c}{CNe+RNe} \\ 
 & N &  $\bar{\Delta M_V}(t_2)$  &  $\bar{\Delta M_V}(t_3)$  & N &  $\bar{\Delta M_V}(t_2)$  &  $\bar{\Delta M_V}(t_3)$ \\ 
 \hline
ALL & 50 & $0.20\pm0.14$  & $0.25\pm0.14$  & 65 & $0.34\pm0.17$  & $0.39\pm0.18$ \\ 
\hline
LF & 13 & $0.43\pm0.27$  & $0.59\pm0.28$  & -  & $0.43\pm0.27$  & $0.59\pm0.28$ \\ 
MF & 10 & $0.82\pm0.33$  & $1.01\pm0.35$  & 15 & $1.07\pm0.24$  & $0.98\pm0.25$ \\ 
HF & -  & -  & -  & 10 & $1.24\pm0.78$  & $1.57\pm0.78$ \\ 
\hline
S & 8 & $0.26\pm0.37$  & $0.27\pm0.36$  & 11 & $0.25\pm0.30$  & $0.36\pm0.31$ \\ 
P & 5 & $-0.89\pm0.37$  & $-0.81\pm0.38$  & 11 & $0.15\pm0.69$  & $-0.22\pm0.73$ \\ 
D & 12 & $-0.01\pm0.24$  & $0.23\pm0.25$  & -  & -  & -\\ 
C & -  & -  & -  & 1 & $4.49$  & $3.88$ \\ 
O & 4 & $0.62\pm0.22$ & $0.56\pm0.20$ & -  & -  & -\\ 
F & 1 & $2.41$  & $2.38$  & - & -  & - \\ 
J & 6 & $-0.52\pm0.31$  & $-0.62\pm0.28$  & -  & -  & -\\ 
\hline
He/N & 4 & $0.34\pm0.53$  & $0.12\pm0.52$  & 9 & $1.64\pm0.49$  & $1.11\pm0.45$ \\ 
Fe II & 34 & $0.16\pm0.16$  & $0.17\pm0.17$  & -  & -  & - \\ 
Hyb & 5 & $-0.37\pm0.19$ & $-0.30\pm0.17$ & 7 & $-0.30\pm0.14$ & $-0.30\pm0.13$ \\
\hline
\end{tabular}
\end{table}

\section{Novae in the Galaxy} 
\label{SecSpatial}
In this section, the statistical results on decline times, absolute magnitudes at  
outburst maximum, and the spatial distribution along with Galactic scale height and space 
density were investigated. Using the results obtained in this section, we also estimated the nova rate for disk novae. 

In further analyses, we divided the sample into two sub-samples corresponding to the used calculation method of the distance and/or absolute magnitude at maximum outburst.
\begin{itemize}
\item Sample 1: Only the novae with reliable distance measurements as listed in Table \ref{table1} were used.
\item Sample 2: For the novae in this sample, only the MMRD relations for $t_{3}$ following $t_{2}$ times were used, assuming that the MMRD relation derived in Section \ref{SecMMRD} is correct.
\begin{enumerate}
\item[2a:] The novae in this sample have reliable light curve parameters.
\item[2b:] The novae in this sample have unreliable light curve parameters
\end{enumerate}
\end{itemize}
Note that, we did not use the MMRD relations for 22 RN and cRN (including likely RN 
systems with high FWHM H$_{\alpha}$>3500 km/s) owing to deviations from the MMRD relations.

\subsection{Decline Times and Absolute Magnitudes}
\label{sec:dec_times_abs}
\begin{figure} \centering
\includegraphics[width=.35\textwidth]{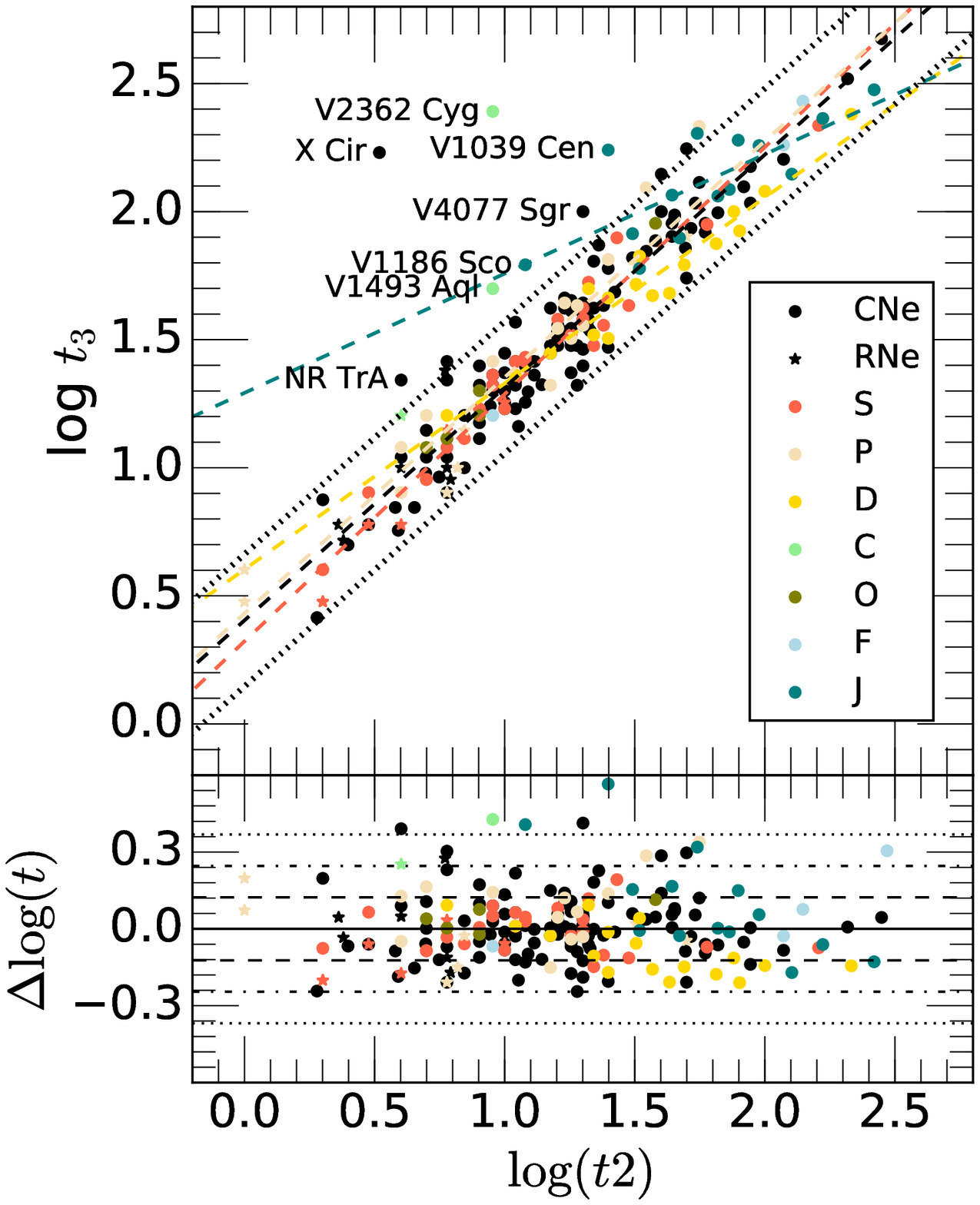} \caption{The distributions of $t_2$ vs $t_3$ decline times. Dashed lines represent the relations between decline times corresponding to subsamples by light curve shapes. The dotted lines are for the $\pm 2 \sigma$ ranges from the relation obtained for all novae in the sample.}
\label{t2_vs_t3} \end{figure}

The decline from outburst maximum is generally smooth, 38\% of the novae \citep{2010AJ....140...34S}, but a considerable number of novae shows jitters, oscillations, dips or flares in their light-curve.
This heterogeneity in nova light curve shapes is a result of the variety of nova outbursts, and it may cause a non-linear decline even at relatively stable decline times ($t_2$ and $t_3$).  
The non-linear decline of the light curve also affects calculations of absolute magnitudes from MMRD relations. Because of these reasons,
we compared decline rates by considering the light
curve shape classification \citep{2010AJ....140...34S} in Fig. \ref{t2_vs_t3}. The median of differences between logarithmic decline times, and between absolute magnitudes calculated from MMRD relations for $t_2$ and $t_3$ times are listed in Table \ref{tab:decline_times_table}.
From the distribution of decline times, a relation
between $t_2$ and $t_3$ times were obtained 
as $\text{log} t_3= 0.91 (\pm0.02) \times
\text{log} t_2 + 0.41 (\pm0.02)$ with $R^2=0.93$ 
and one standard deviation error of $\pm 0.12$ day. Our relation between decline times is consistent with that found by \citet{2003cvs..book.....W}, $\text{log} t_3= 0.88 \times
\text{log} t_2 + 0.44$. Here, the intercept value of 0.91 implies that the novae have nearly linear declines between the $t_2$ and $t_3$ times.
However, 12 of 192 novae have larger separations between the logarithmic $t_2$ and $t_3$ times (Fig \ref{t2_vs_t3}), especially for X Cir, V2362 Cyg and V1039 Cen. For these scattered novae, the differences between absolute magnitudes from MMRD relations for $t_2$ and $t_3$ are larger than $\pm2\sigma\sim 0.5$ mag. Note that these systems were excluded from the samples for spatial distribution analyses. On the other hands, we also obtained relations between decline times for subsamples of light curve shape classifications, which have more than 5 systems, as following; 
\begin{itemize}
\item For S-types: $\log t_3=0.96 \log t_2 + 0.32$, $R^2=0.95$, $\sigma=0.04$
\item For P-types: $\log t_3=0.92 \log t_2 + 0.43$, $R^2=0.90$, $\sigma=0.07$
\item For D-types: $\log t_3=0.72 \log t_2 + 0.60$, $R^2=0.93$, $\sigma=0.05$
\item For J-types: $\log t_3=0.46 \log t_2 + 1.29$, $R^2=0.57$, $\sigma=0.12$
\end{itemize}
The relations for S- and P-types are very similar to those obtained from all systems, but the novae having D- and J-types seem to follow different trends that indicate these systems may not have non-linear declines. However, these scattered novae cover only $1\%$ of the whole sample, and  the mean of differences between absolute magnitudes (Table \ref{tab:decline_times_table}) are the within limits of $\pm2\sigma=0.5$ mag. Thus, we assumed this effect is negligible in the calculation of the absolute magnitudes from MMRD relations for the $t_2$ and $t_3$ times.

\begin{table}
\caption{Comparison of logarithmic decline times, and absolute magnitudes. N is number of novae in the sample, $\Delta t$ is residuals, $\log(t_3)-\log(t_{3,c}))$, $\Delta M_V(t_2,t_3)$ is median of differences between absolute magnitudes obtained from MMRD relations for $t_2$ and $t_3$ times.}
\label{tab:decline_times_table}
\begin{tabular}{l@{\hskip3pt\vline\hskip3pt}ccc@{\hskip3pt\vline\hskip3pt}cc}
\hline
 Sample & & CNe & & & CNe+RNe  \\
  & N & $\Delta t$ & $\Delta M_V(t_2,t_3)$ & N & $\Delta t$ \\
 \hline
All  & 192 & $-0.0\pm0.01$ & $0.06\pm0.03$  & 214 & $0.0\pm0.01$\\ 
S & 26 & $0.00\pm0.02$ & $0.04\pm0.04$  & 32 & $-0.03\pm0.02$\\ 
P & 13 & $0.12\pm0.04$ & $0.32\pm0.08$  & 21 & $0.05\pm0.03$\\ 
D & 17 & $-0.11\pm0.02$ & $-0.11\pm0.02$  & - & -\\
C & 2 & $0.77\pm0.24$ & $1.82\pm0.56$  & 3 & $0.43\pm0.22$\\ 
O & 5 & $0.04\pm0.02$ & $0.13\pm0.05$  & - & -\\
F & 4 & $0.02\pm0.07$ & $0.15\pm0.17$  & 4 & $0.02\pm0.07$\\ 
J & 14 & $0.03\pm0.05$ & $0.15\pm0.12$  & - & -\\
\hline
\end{tabular}
\end{table}
The absolute magnitude of the Galactic novae, excluding RN+cRN, typically range from $M_{V_{max}}\simeq-5$ to -10 mag. For the novae in sample 2, the histogram distribution could be fitted with a Gaussian distribution with a  peak at $M_{V_{max}}\simeq -7.9\pm0.9$ mag. Even though, when the novae only with reliable light curve (sample 2a) are considered, we get the same result for the Gaussian peak.
This value is consistent with \citet{2009ApJ...690.1148S}  where the mean 
peak absolute magnitude for Galactic novae was obtained to be -7.8 mag, but smaller than  their estimate of -7.2 mag for M31. 
For the novae in sample 1, the mean absolute magnitude changed to $M_{V_{max}}\simeq 
-7.2\pm0.24$ mag, which is consistent with that adopted by \citet{2017ApJ...834..196S}, but the 
histogram distribution is broad without a significant peak. 
The maximum magnitudes mainly depend on the discovery magnitudes so that the absolute 
magnitude at maximum is likely underestimated. Interstellar extinction obscure relatively faint Galactic novae,  
so the mean absolute magnitude at outburst maximum for the Galaxy could be biased  towards 
brighter novae, corresponding to the Galactic disk population in Solar neighbourhood. In contrast,  
as with the case of both novae V598
Pup and KT Eri \citep{2010ApJ...724..480H}, the brightest novae that can even be seen by naked eye, may not be discovered unless they are monitored in different phases such as X-ray, and 
be missed by conventional ground-based observing techniques. Thus the Galactic (and extra galactic) nova sample is not free from selection effects that likely bias the results.

\begin{figure} \centering
\includegraphics[width=0.48\textwidth]{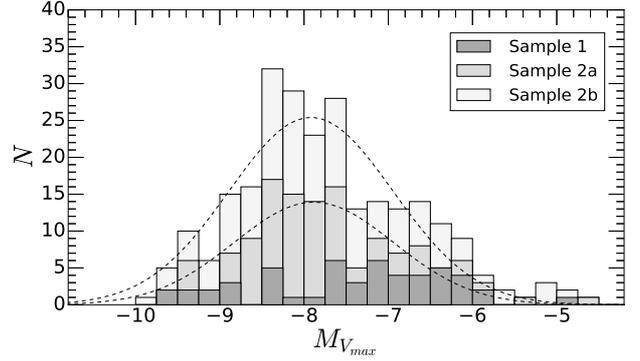} \caption{Absolute magnitude distribution of the Galactic nova samples. The distributions are fitted by Gaussian functions which are represented by dashed lines.}
\label{fig:hist_MV} \end {figure}

\subsection{Spatial Distributions}
\begin{figure} \centering
\includegraphics[width=0.48\textwidth]{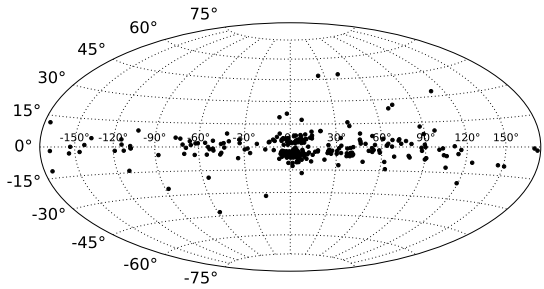} 
\caption{Galactic coordinates of novae in our sample.}
\label{fig:fig3} \end {figure}

The Galactic distribution of all Galactic novae in our catalogue is given
in Fig. \ref{fig:fig3}. In this distribution, systems are
mainly located at low Galactic latitudes ($b\leq 10 \degrees$)
especially towards the Galactic bulge. 
Since the Galactic novae in the sample concentrate in the Galactic plane, where the bulk 
of the absorbing medium is located, the extinction for many of them can be high and can 
not be neglected. 
Although, interstellar reddening estimates were given for nearly half of 
the novae in the catalogue, the extinction and distances for the remaining 
novae were simultaneously determined by an iteration method: We first 
obtained the interstellar extinction for  each distance step in the line of 
sight following \citet{2011ApJ...730....3S} and 
\citet{2014MNRAS.437..351B}. We then calculated the absolute magnitude of 
each nova from the MMRD relation. Using the apparent magnitude and the absolute 
magnitude we can calculate the corresponding possible distance extinction 
values for each nova. We estimated the distance and the extinction of each 
nova simultaneously by calculating the point where the distance and the 
extinction derived from the distance modulus ($V_{max}-
M_{V_{max}}=5\log$d$-5+A_V$) matches the values derived from the dust 
extinction model. To test the robustness of our method we also followed 
this method for the novae which already have independent extinction 
measurements. The standard deviation of the differences between the 
reddening derived from the method summarized above to the independent 
measurements is found to be $\pm 0.08$ mag. We used this value as our error 
on determining the interstellar reddening towards the sources which have no 
independent reddening measurements. Using this iteration method with the 
MMRD relation only for the novae without reddening estimates in sample 2 
provides us a large sample to investigate the spatial distribution and 
Galactic model parameters of Galactic novae.

\begin{figure} \centering
\includegraphics[width=0.3\textwidth]{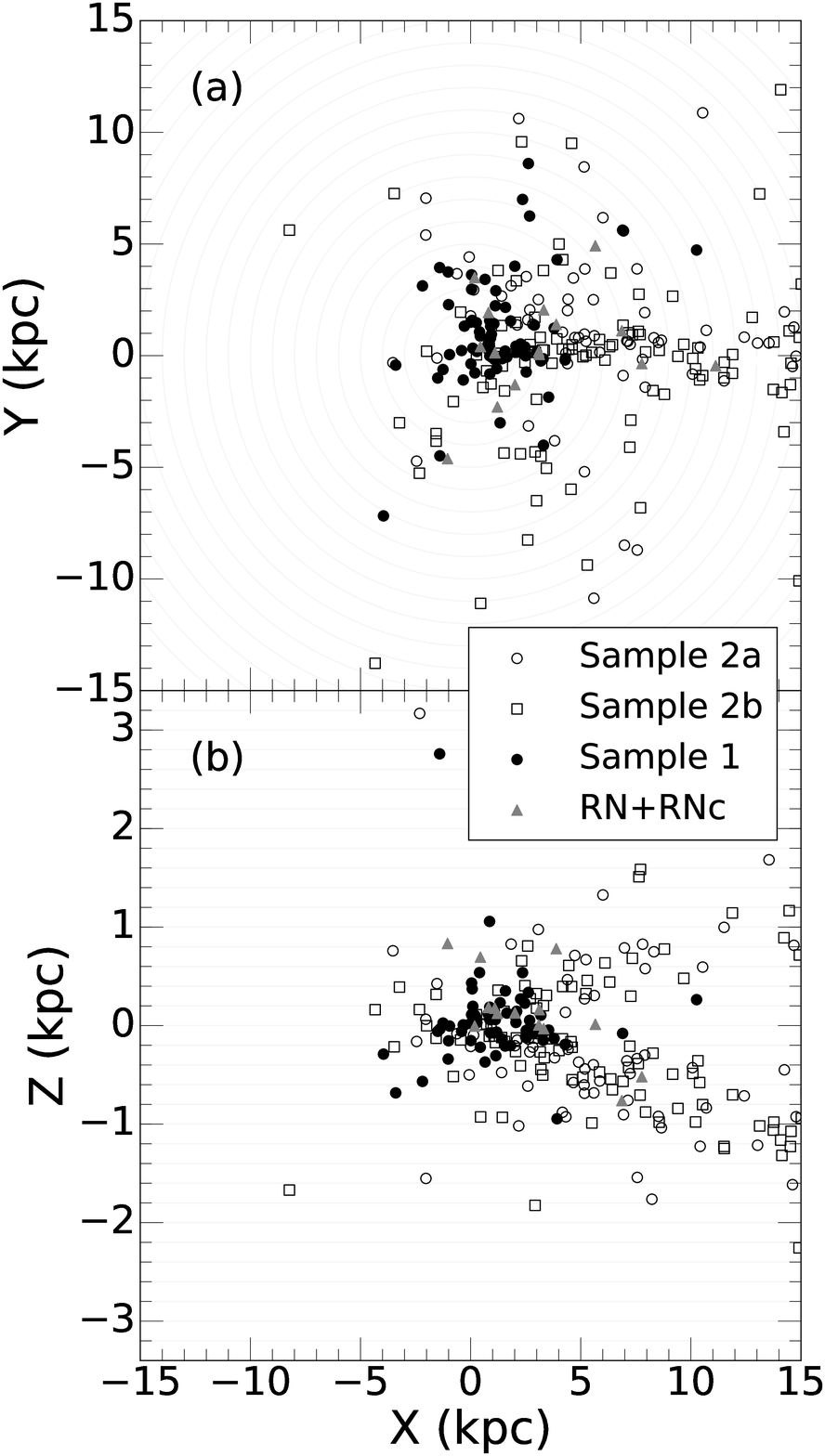} 
\caption{The projected positions of novae in our sample (a) on the X-Y Galactic plane and (b) on X-Z plane perpendicular to it. 
}
\label{fig:fXYZ} 
\end {figure}

To inspect the Galactic distribution of Galactic novae, heliocentric distances (X towards Galactic centre, Y towards the Galactic rotation, Z towards the Galactic north Pole) are obtained, and the projected positions on the Galactic plane (XY) and on the plane perpendicular to it (XZ) are derived (Fig.\ref{fig:fXYZ}). 
The Galactic novae in our sample are mainly concentrated in the Galactic disk (Fig. \ref{fig:fXYZ}b) and towards the inner Galaxy (Fig. \ref{fig:fXYZ}a), but the novae in the neighborhood of the Galactic center are located at the edge of bulge. The novae in sample 1 occupy a smaller spatial volume than the novae in sample 2 where a number of novae (11 of 120) with uncertain light curve novae in sample 2  has unrealistic distances (>20 kpc) even beyond the Galactic bulge through Galactic disk. To avoid such unrealistic measurements, we set an upper limit on the distance calculations from MMRD relations of 20 kpc. 
The median values of XYZ distances for sample 2 are 4.30, 0.39 and $-0.13$ kpc, while the novae in sample 1 have about 1.0, 0.5, and 0.02 kpc, and these do not change even with adding RNe.  
The numbers of systems with $(X, Y, Z) \geq 0$ are
(52, 43, 30 of 65) for sample 1, (66, 53, 46 of 80) for sample 1 when adding RNe, and (218, 156, 100 of 249) for sample 2. In short,
$\sim85\%$, $\sim65\%$ and $\sim50\%$ of Galactic novae in samples have Galactic distances of $(X, Y, Z) \geq 0$, respectively.
The results indicate that there is a strong bias towards the Galactic bulge. 
Since the sky surveys are mainly concentrated towards 
the Galactic bulge, this bias probably arises from observational selection effects. 
Although, the number of the bulge novae is expected to be more than half of all novae as in 
M31 \citep{2006MNRAS.369..257D}, the Galactic novae in our sample are mainly located on the edge 
of the Galactic bulge likely due to increasing interstellar reddening effect towards the 
Galactic center. This distribution indicates that the bulge novae could not be observed in 
our sample.

\begin{figure} \centering
\includegraphics[width=0.4\textwidth]{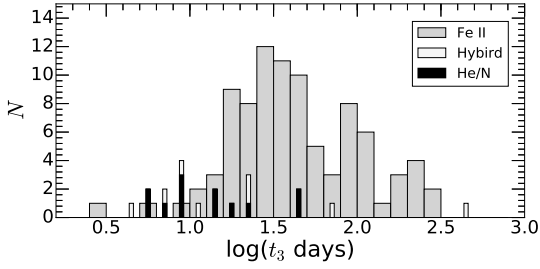} 
\caption{The histogram distributions of decline times for the Galactic novae corresponding to their spectral classes.}
\label{fig:t3_hist} 
\end {figure}

\begin{figure} \centering
\includegraphics[width=0.49\textwidth]{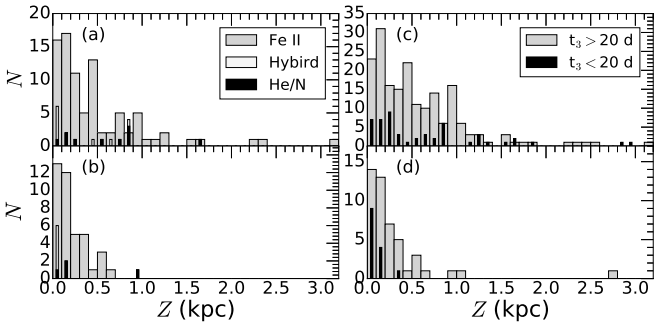} 
\caption{Z-distance histogram distributions  of the Galactic novae corresponding to their (a, b) spectral and (c,d) speed classes, (a,c) for the novae in sample 2, (b,d) for the novae in sample 1.}
\label{fig:Zspec} 
\end {figure}

In the studies of the spatial distribution of the novae, the thin disk population is characterized by bright and fast novae, while slow and faint novae belong to the thick disk or bulge population.
The differences between the two populations are explained by the different nature of the nova progenitors \citep{1992A&A...266..232D}. 
Fast novae characterized as He/N novae are believed to be associated to relatively massive WDs, while slow novae related to Fe II novae have less massive WDs. Since more massive WD requires smaller mass of the accreted material to start thermonuclear runaway, the more violent outburst theoretically occurs with higher expansion velocity and fast decline. Hence, these novae with fast decline time related to He/N evolve more quickly, and they should be located closer to the Galactic plane.
\citet{1998ApJ...506..818D} proposed that fast novae ($t_3< 20$ day) belonging to the He/N 
spectroscopic class are preferentially concentrated through Galactic plane ($Z<$ 100 pc)
related to Pop I, while Fe II novae with slow decline time ($t_3 > 20$ day) 
are located up to $>1000$ pc from the Galactic plane, and they are likely related to 
the thick disk/bulge (Pop II) population.  
To test this prediction, we considered the histogram distribution of these systems based on 
their spectral and speed class. In the samples, $\sim80\%$ of Fe II novae (72 of 87) are slow novae 
($t_3 \geq 20$ days), while the percentage of fast novae with $t_3 < 20$ days belonging to the
He/N+hybrid class  is $\sim60\%$ (11 of 18)). Note that we considered the hybrid class sources similar to He/N novae for the analyses.
Fig. \ref{fig:t3_hist} presents the histogram distribution of the $t_3$ times 
according to the spectral classification, where both Fe II and He/N+hybrid novae are scattered throughout the 
whole range of $t_3$ times, but the peak values of the histogram distribution for Fe II and He/N+hybrid 
are at $t_3=$37 and 13~days, respectively. 
The vertical Z-distances for  their spectral classes (Fig. \ref{fig:Zspec}a) 
indicate that vertical distances for Fe II and He/N+hybrid novae in sample 2 extend up to $\sim3.0$ and $\sim1.7$~kpc 
from the Galactic plane with median values of vertical Z-distances of 0.3 and 0.33~kpc,
respectively. Nine of Fe II novae ($10\%$) were found higher than 1000 pc above the plane, and 16 of Fe II novae ($18\%$) are located in the Galactic plane ($Z<100$ pc). For the He/N+Hybrid novae, the number of systems with $Z>1000$ pc is only 1 ($6\%$), and 6 of them ($33\%$) are at $Z<100$ pc. 
The median Z-distances changes to 0.15 and 0.03 kpc for the novae in sample 1 (0.15 and 0.12 kpc with adding RNe), but the novae classified as hybrid class in both sample 1 and 2 are at shorter vertical Z-distances. Since most of the novae in sample 1 are nearby (Fig.\ref{fig:fXYZ}) as mentioned above, their Z-distances extend only up to 1 kpc.
The distributions of Z distances according to fast and slow novae show similar trends as derived for the spectral classification. Moreover, both slow and fast novae in sample 2 (Fig \ref{fig:Zspec}) are located up to $\sim4$ kpc from the Galactic plane. Large part of fast novae (11 of 51, $22\%$) and only 25 of 189 slow novae ($13\%$) have vertical distances from the Galactic plane higher than 1000 pc, while only 7 of 51 fast novae ($14\%$) and 23 of 189 slow novae ($12\%$) lie in the Galactic plane at $Z<100 pc$.
These trends do not change even for the novae in sample 1 when adding RN+cRN.
Although, \citet{1998ApJ...506..818D} concluded that spectral or speed classification correlates 
with the stellar population. Such relations were not derived from our 
samples which contain much more novae with spectral classification. On the contrary,
both Fe II and He/N novae or fast and slow novae 
are distributed in the whole range of vertical distances, but are mainly 
concentrated throughout the Galactic disk. However, a similar selection bias (as mentioned in section \ref{sec:dec_times_abs}) may effect this result as well. The Galactic nova samples seem to contain the systems, which are located in the 
Galactic disk structure related to Pop I or young Pop II. The faint/undiscovered novae belongs to thick disk or halo components may show a correlation with spectral or speed classifications.

\subsection{Galactic Model Parameters of the Galactic Novae}
\label{SecGMP}
In this section, we investigate the Galactic model parameters of the Galactic 
novae such as the scale height and space density. Our sample contains novae erupted as far back as the 19th century resulting in a sample covering $\sim 200$~yr of 
observations, but  the systematic and effective discoveries of novae have only been possible since 1960s so the effective observing time of roughly 50 years is used in the following calculations. Since only the novae with observed outbursts are included in the sample, there is a completeness problem arising from unrecorded nova outbursts due to all kind of observational effects. This incompleteness hardly influences the vertical spatial (Z) distribution but strongly affects the space density as mentioned in previous studies \citep{1984ApJS...54..443P, 1984Ap&SS..99..363D,1993A&A...275..239D,2017ApJ...834..196S}. 
Thus, the space density $\rho_0$ can only be calculated from the observed local outburst density $\rho_{out}(0)$ for the discovered classical nova sample by taking into account the recurrence time ($T_R$); $\rho_0=\rho_{out}(0)\times T_R$. A detailed discussion of both the space density and the recurrence time calculation in previous studies were given in \citet{1993A&A...275..239D}, which was the last study on the space density of the Galactic nova population. 

The recurrence time of classical novae, which may range from a few hundred to thousand years, depends on several factors, for example the mass of the WD, the donor star, the amount of ejected material, or the rate of mass transfer. It even changes with the stellar population and the type of the host galaxy \citep{2016MNRAS.458.2916C}. However, the mean recurrence times for classical novae were estimated to be between $T_R=3000-26000$ years \citep{1984ApJS...54..443P, 1984Ap&SS..99..363D, 1993A&A...275..239D}.
Since the calculated recurrence times show a very large scatter, the space density determined in this way will be only a limit for the actual value. Thus, we adopted two recurrence times as 3000 and 26,000 years to set a lower and upper boundary on the space density for classical novae.

On the other hand, in order to be able to determine the Galactic model parameters, the Galactic novae should be separated into samples based on stellar populations. Although not very effective, the only way to estimate the stellar population is to deduce Z-distances. In our sample,  82\% of the novae have vertical distances smaller than 0.825 kpc. Note that as discussed below, for the sake of completeness we only used the novae with distances up to 4~kpc in our further estimations, and 99\% of these  novae have Z-distances smaller than this limit. 
Here, the vertical distance of 0.825 kpc is important, since the spatial densities of thin and thick disks are equal at this mode value of the vertical distance \citep{2015Ap&SS.357...72A}.
Thus, almost all Galactic novae in the sample can be considered as 
members of the thin-disk component of the Galaxy.
This is a reasonable assumption
because the population analysis already showed that $94\%$ of
cataclysmic variables in the Solar neighbourhood belong to the thin-disk
component of the Galaxy \citep{2015Ap&SS.357...72A}. The contribution of thick disk or halo CVs to the thin disk sample adds a small effect (less than $4\%$) to the scale height 
estimations \citet{2015NewA...34..234O}. In this case, the effect of contribution can be considered as 
negligible. Thus, Galactic novae in our sample were not classified based on the 
population types in the estimation of the Galactic model parameters and all are assumed to be thin disk members. 

\begin{figure*} \centering
\includegraphics[width=0.80\textwidth]{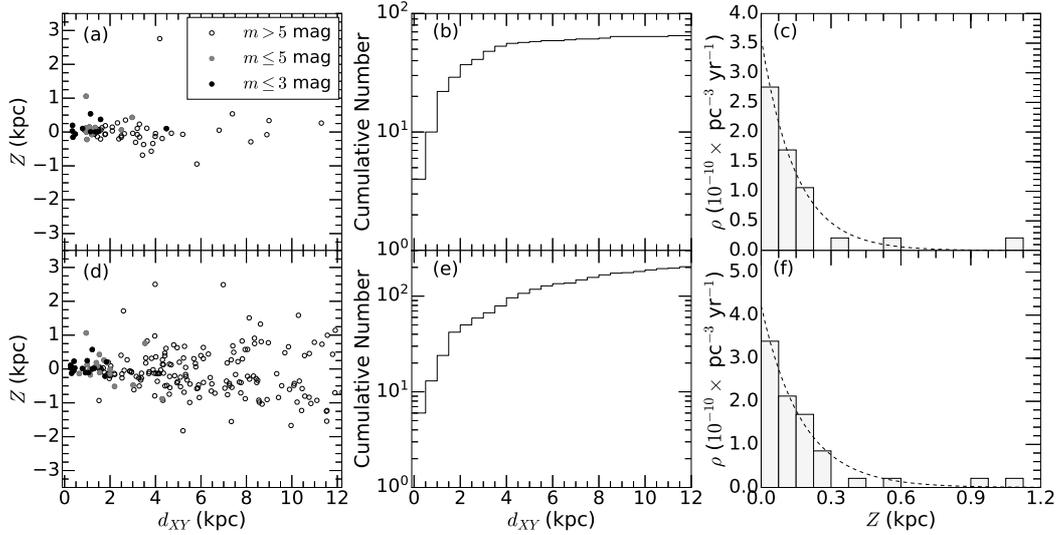} 
\caption{(a,d) Z-distance of the Galactic novae in the XY direction, (b, e) Cumulative numbers of the Galactic novae occupying a cylindrical volume of with radius $d_{XY}$, (c, f) the outburst density in the z-direction within a cylinder with a radius of $d_{XY}=2$ kpc, (a, b, c) for CNe in sample 1, (d, e, f) for CNe in sample 2. }
\label{fig:g_z} 
\end {figure*}

The local outburst density can be obtained using the Galactic novae within a cylinder 
centered at the Sun with a radius and infinite height.
 \citet{1984Ap&SS..99..363D} used a cylinder radii of 
$d_{XY}=1000$~pc and assumed that all the novae in this region are already known. This is a reasonable assumption, since the completeness of the novae decreases after this radius where the novae at the edge 
reach to an apparent magnitude of 4 to 5~mag. In Figure \ref{fig:g_z}a,d, the distribution of the z-distances 
of the novae in our sample are given as a function of their $d_{XY}$ distances. 
This distribution implies that the novae with apparent magnitude brighter than 3 and 5~mag reach to $d_{XY}=2$ and 4~kpc, but most of them are concentrated within 2~kpc. 
The cumulative number distribution of the novae in different cylinders with radii $d_{XY}$ 
are given in Fig. \ref{fig:g_z}b,e. The number of the discovered novae should 
increase with larger radius, but this trend slows after 2~kpc, and it becomes almost 
constant after $\sim 4$~kpc. Since the discovery efficiency has increased 
in last decades, it is possible to assume a larger radii than that adopted by 
\citet{1984Ap&SS..99..363D} within which all the possible classical novae have already been discovered. 
From the classical novae  sample 1 and 2 within the cylinder with $d_{XY}=2$~kpc, the outburst space density in the solar neighborhood is calculated as 
 $\rho_{out}(0)=3.57 \pm 0.25$ and $\rho_{out}(0)=4.22 \pm 0.25 \times \ 10^{-10}$ pc$^{-3}$ yr$^{-1}$ with a scale 
height of 148~pc and 175~pc (see Figure \ref{fig:g_z}c,f), respectively. The results of the calculations are given in Table \ref{tab:gmp_result_table}. If 4~kpc is adopted as the radius of the cylinder, the outburst density 
can be found to be as $1.3$ and $1.47 \times \ 10^{-10}$ pc$^{-3}$ yr$^{-1}$ with a scale height of $\sim165$ pc and $\sim209$ pc for the classical novae in sample 1 and 2,
respectively. Using 1 kpc radius, we calculated $\rho_{out}=6.05 \times \ 10^{-10}$ pc$^{-3}$ yr$^{-1}$ for only the novae in sample 2.
In our analyses, the radii of the 
cylinder and/or using different samples, depending on reliable distance measurements (sample 1) or the MMRD relation (sample 2), did not significantly change the local outburst density.
So even if a MMRD relation is not realized, our result is consistent.
It is in the range of $1.3-6 \times 10^{-10}$ pc$^{-3}$ yr$^{-1}$. Our result is 
slightly larger than previous observational estimates, cf. 
$1.7 \times \ 10^{-10}$ \citep{1984ApJS...54..443P}, $3.8 \times \ 10^{-10}$  \citep{1984Ap&SS..99..363D}, 
1 -- 4.4 $\times \ 10^{-10}$\citep{1992MNRAS.258..449N},  
$0.30 \times \ 10^{-10}$ pc$^{-3}$ yr$^{-1}$ \citep{1993A&A...275..239D}, as well as the semi-model estimate of $\sim1 \times \ 10^{-10}$ pc$^{-3}$ yr$^{-1}$ for disk novae \citep{2017ApJ...834..196S}.
Using $T_R=3000$ and 26,000~yr, the space density is calculated as 
1.07 $\times 10^{-6}$ and 9.3 $\times 10^{-6}$ pc$^{-3}$ for the classical novae in sample 1 within the cylinder with a radius of $d_{XY}=2$ kpc, while for the novae in sample 2, it is calculated as 1.3 $\times 10^{-6}$ and 11.0 $\times 10^{-6}$ pc$^{-3}$. Taking into account all calculations for all samples and cylinder radii,
the range of the space density of novae is obtained as 0.4~--~15.7  $\times 10^{-6}$ pc$^{-3}$.

\begin{table}
\caption{Result for spatial distribution analyses, $d_{XY}$ is radius of cylinder, $\rho_{out}$ is the local outburst space density, H is the scale height, $\rho_0$ is the local space density using $T_R=3000$ and 23000 yr$^{-1}$.}
\label{tab:gmp_result_table}
\begin{tabular}{l@{\hskip4pt}c@{\hskip4pt}c@{\hskip4pt}c@{\hskip4pt}c@{\hskip4pt}c@{\hskip4pt}c}
\hline
 & $d_{XY}$ & $\rho_{out}$ & $H$ & \multicolumn{2}{c}{$\rho_0$} & $N_{rate}$\\ 
  &  (\scriptsize{kpc}) & (\scriptsize{$\times 10^{-10}$ pc$^{-3}$ yr$^{-1}$}) &  (\scriptsize{pc}) & \multicolumn{2}{c}{(\scriptsize{$\times 10^{-6}$ pc$^{-3}$ yr$^{-1}$})} &  (yr$^{-1}$)\\ 
 &  &  &  & \scriptsize{$T_R=3000$} & \scriptsize{$T_R=26000$} &  \\ 
 \hline
Sample 1 & 2 & $3.57\pm0.25$ & $148\pm17$ & 1.07 & 9.3 & $54^{+16}_{-13}$\\ 
 & 1 & -- & -- & -- & -- & --\\ 
 & 4 & $1.30\pm0.13$ & $165\pm21$ & 0.39 & 3.4 & $23^{+9}_{-7}$ \\ 
 \hline
Sample 2 & 2 & $4.22\pm0.26$ & $175\pm17$ & 1.3 & 11.0 & $80^{+21}_{-16}$\\ 
 & 1 & $6.05\pm1.46$ & $147\pm62$ & 1.8 & 15.7 & $95^{+92}_{-65}$\\ 
 & 4 & $1.47\pm0.11$ & $209\pm22$ & 0.4 & 3.8 & $35^{+9}_{-8}$\\
\hline
\end{tabular}
\end{table}

Computing the local outburst density of Galactic novae corresponding to the 
Galactic disk allows us to estimate the disk nova rate. For calculations, 
we used the same assumptions and direct-extrapolation method as described 
in \citet{2017ApJ...834..196S}, who calculated the overall Galactic nova rate by extrapolating the nova rate (assuming different observed rate for novae reaching $m \leq 2$) 
for a given apparent magnitude to include the entire Galaxy based on  
Bahcall \& Soneira (1980)'s models of bulge and disk distributions. 
Since we calculated the local outburst density and the scale height observationally, 
we only changed these parameters as given in table \ref{tab:gmp_result_table}, 
while we adopted mean values of absolute magnitude of 
$-7.2 \pm 0.24$ and $-7.9\pm0.9 $ mag as calculated for sample 1 and 2, respectively, 
in section \ref{sec:dec_times_abs}. The results are given in Table \ref{tab:gmp_result_table} that shows disk nova rates are in the range of $\sim20$ to $\sim100$ per year. An average of the calculated plausible parameters, $\rho_{out}= 3.9\pm0.25 \ \times 10^{-10}$ pc$^{-3}$ yr$^{-1}$, $H = 160$ pc, $ M_{Vmax} = -7.5\pm0.9$ mag, yields a disk nova rate of $N_{rate}= 67^{+21}_{-17}$ yr$^{-1}$. Our results are slightly larger than the results adopted by  \citet{2017ApJ...834..196S}. They calculated disk nova rates in the range of $\sim35$ to $\sim110$ per year by assuming different observed rate of the novae with apparent magnitude less than 2 mag and different mean absolute magnitudes (similar values as in our study) for Galactic disk novae. They adopted $50^{+31}_{-23}$ yr$^{-1}$ as a best estimate for an average of global nova rates (bulge + disk), where the disk nova rate was calculated as $45^{+27}_{-21}$ per year. Our average estimate is slightly larger than the average estimate for the disk nova rate in \citet{2017ApJ...834..196S}, but it is consistent within errors. We should note that the relatively larger nova rate calculated here is probably arising from using a larger local outburst density. However, if the rate of undiscovered novae is higher than expected, both local outburst density and the nova rate would be increase to larger values. For example,  our result would be consistent with \citet{2017ApJ...834..196S}, if observed rate of novae (with $m\leq2$) assumed between $43\%$-$90\%$. 

\section{Results}

In this study, we investigated the MMRD relation and spatial distributions of Galactic novae 
from a new compiled catalogue. The catalogue, which will be updated frequently, has fundamental 
parameters, but more importantly it contains the reddening and
distance estimates from the improved RDRs.
We continue to collect the related parameters, and study the Galactic novae spectroscopically
to obtain their interstellar reddening/extinctions that allow us to
limit their distances, and investigate their spectroscopic features. In addition, the distances in GAIA DR2 will be added to the catalogue.
Hence, this catalogue may help to resolve uncertain distances of some nova systems, 
which have lot of variety. By this way, the nature of the novae
in the Galaxy might be clarified in the future. In our (spatial) analyses, we used two different samples. The first sample contains the novae whose distance measurements (non-MMRD) exist, while we only used MMRD relations for the second sample. Main results of our paper can be summarized as follows.
\begin{itemize}
\item Using the distances in the catalogue, we obtained MMRD relations for $t_2$ and $t_3$ decline times with an scatter of $\sim 1$ mag. We suggest that there is a possible dependence of light curve shapes and FWHM H$_\alpha$ on MMRD relations. 
\item The results obtained by using the two different samples for the spatial analyses  are consistent with each other. Thus,  we confirm that the MMRD relation is still useful tool for statistical analyses.
\item The distributions of decline times $t_2$ and $t_3$ indicate that there is a relation between decline times, but it depends on light curve shape classifications, especially for D- and J-types.
\item From the histogram distributions of mean absolute magnitude at outburst maximum, we obtained $M_{V_{max}}=-7.2\pm0.24$ and $-7.9\pm0.9$ mag for the samples.
\item The spatial distributions of discovered Galactic novae imply that the systems 
are mainly placed in the Galactic disk towards the Galactic bulge, but  belong to the disk population. Besides, we did not found any correlation of the stellar population with spectral or speed classification.
\item For the calculation of the Galactic model parameters, we used the novae within cylinders with various radii.
The best estimates for the local outburst density are calculated as 3.6 and 4.2 $\times 10^{-10}$ pc$^{-3}$ yr$^{-1}$ with a scale height of 148 and 175 pc for two different samples, while a possible range is $1.3-6 \times 10^{-10}$ pc$^{-3}$ yr$^{-1}$. Using $T_R=3000-26000$ yr,
the most plausible range for the local space density is calculated as 1-10 $\times 10^{-6}$ pc$^{-3}$.
\item A simple calculation on disk nova rate suggest that it is in range of $\sim20$ to $\sim 100$ yr$^{-1}$ with an average estimate $67^{+21}_{-17}$.
\end{itemize}

\section{Discussions and Conclusions}
\label{SecDisCon}

Based on a new Galactic nova catalogue, the MMRD relation and the spatial distribution of novae are investigated. The MMRD
relation is an effective tool (in fact it is the only one) 
when it is applied in a statistical way to a large sample, 
but it is not very reliable for individual novae
because of the large individual errors and scatter in the overall relation. 
For a single nova, solely relying on this relation, 
the absolute magnitude may differ $\sim1$~mag from the actual value. 
In order to consider any possible systematic effect on the MMRD relations due to the distance measurement method, 
MMRD relations for three different samples distinguished by distance measurements methods were obtained, but we did not find any significant bias, and we adopted the MMRD relation that depends on the novae whose distances calculated from two different methods (parallax and RDRs).

Recent studies of \citet{2011ApJ...735...94K} and \citet{2017ApJ...839..109S}
claimed that a new class of faint, fast and non-recurrent novae ($t_3<20$ days) exists in
M31 and M87. They adopted them as non-recurrent novae since the majority of them have low velocities (FWHM < 2500 km/s) and belong to Fe II class. These novae are up to 3 magnitudes fainter than
predicted by the MMRD relation in \citet{2000AJ....120.2007D}. 
These studies caused a decrease in reliability of MMRD relations.
This point should be further investigated to clarify the problems of the MMRD distributions, or to understand these types of novae.
For example, \citet{2017MNRAS.469.4341M} commented that none of the other surveys detected them during the whole observational history for M31 \citep[e.g.][]{2006MNRAS.369..257D}.
Moreover the OGLE
microlensing survey has continuously monitored the Magellanic
Clouds, where reddening is not an issue, no novae of the type claimed by \citet{2011ApJ...735...94K} were detected \citep{2016ApJS..222....9M}. 
\citeauthor{2017MNRAS.469.4341M} (\citeyear{2014ASPC..490..183M,2017MNRAS.469.4341M}) 
suggested that there are systematic effects on \citet{2011ApJ...735...94K}'s analyses owing to noisy and scarcely sampled light curves, frequently detecting the novae while already declining and thus probably missing the true maximum, and underestimated  extinction estimates performed in an unusual way. 
Anyway, there is a possibility that this new type of novae exists in the Galaxy. 
For example, the
novae in Table 1b (V977 Sco, V4332 Sgr, and MU Ser) were not
added in the MMRD diagrams since their light curve parameters or distances/extinctions 
are not well-determined. If we assume their parameters are correct, their absolute magnitudes will be fainter than that calculated from our MMRD.
If the distances/extinctions or light curve parameters are uncertain or
biased, the possibility to verify the existence of an MMRD relation obviously
decreases. Therefore, we only used the novae for which the distances and
the interstellar extinctions were reliably determined. However, there is overall scattering in MMRD distributions similar (but larger) to that observed in \citeauthor{2000AJ....120.2007D}. Generally, this scattering is thought to be due to the distance and extinction, and  Gaia DR2 or any other survey may minimize uncertainties on distance (or extinction) measurements that may decrease scattering on MMRD distributions. However, the scattering would be caused by a second parameter (e.g. spectroscopic type, light-curve type, population type, mass of the underlying WD, chemical composition of the WD or of the accreted matter), which should influence the luminosity at outburst maximum. 
Just the WD mass, the accretion rate/envelope mass and the WD luminosity can produce a rich variety of nova outbursts. The heterogeneity in nova light curves suggests 
that a single parameter may not characterize the decline well. In our study, we suggest that  the MMRD relation likely depends on other parameters in addition to the decline time, as FWHM H$_\alpha$ (an indication of the expansion velocity), the light curve shapes (especially D- and J-types).

Analyses of the MMRD relation considering FWHM H$_\alpha$ show that the scattering on absolute magnitude is increasing with FWHM H$_\alpha$, which is related to the expansion velocity. Moreover, for the RN+cRN with medium velocities, we were able to obtain a relation, whose slope is larger. 
Recurrent novae (or the systems with high expansion velocity) seem to follow a different trend than that obtained for
classical novae. From population models, \citet{2014ApJ...788..164P} suggested that ~25\%
of the systems labeled as CN are in fact RN for which only one
outburst has thus far been discovered. However, we do not actually know the fraction between the classical novae and RNe, and which nova systems are recurrent. There are only 10 known RNe on which the statistical studies depends. Considering the light curve shape, J-type novae have a constant absolute magnitude of -7 mag. The investigation on decline times for all novae in the catalogue indicates that the novae with D- and J- type light curve shapes may not have linear decline. If so, 
only considering the decline times ($t_2$ or $t_3$) may not characterize the absolute magnitude at outburst maximum and the decline well. All of these arise uncertainties when using of MMRD relation if the type of nova is not clarified.
Furthermore, the luminosities of novae may differ depending on their stellar
populations or their evolutionary states; i.e. on the stellar age and metallicity. 
For the scatters found in other distance indicators, e.g. Cepheids, 
it has been suggested that a different metalicity of the host galaxy may significantly affect the zero-point of the relation \citep{2001ApJ...553...47F}. In this case, there is a possibility on any environmental dependence on MMRD relations as well.
Since our and the other studies in literature depend on disk novae in the Galaxy, the contribution of other populations (bulge, halo, even thick disk) is not well known.
Our study does not dispel the concerns regarding the validity the MMRD relation, but has raised open questions. 
To understand the MMRD relation better, the novae in galaxies, which have reliable distances (and extinctions), should be studied 
with high quality photometric and spectroscopic data. A larger sample of them would also be very welcome as well as corresponding parameters and well-determined light curves that allow us to classify their light curve shapes. 
Our catalogue may also help to investigate the nature of MMRD diagrams in the long-term.
Studying the MMRD relation is important to understand the origin
of the nova eruption itself, besides using it as a tool to estimate absolute magnitudes.

In this study, we did not find any signs that the spectral classification represents the stellar population of the Galactic novae, as suggested by \citet{1998ApJ...506..818D}. However, our decline time histogram implies that the He/N+Hybrid novae are fast systems with a peak time at $t_3=13$ days, while the slow Fe II novae have a peak decline time at $t_3=42$ days. In contrast to the suggestion of \citet{1998ApJ...506..818D}, we found that both Fe II and He/N novae or fast and slow novae 
are located in the whole range of vertical distances, but mainly 
concentrated throughout the Galactic disk.
The differences between the results in these two studies may arise from the samples used in the analyses. 
The sample in  \citet{1998ApJ...506..818D} contains a smaller number of novae with spectral classification, which 
reach only to a vertical distance of $\sim2$ kpc, whereas our sample have much more systems reaching up to two times 
farther in vertical direction from the Galactic plane.  
A relatively large number of slow novae located close to the Galactic plane and fast novae farther from Galactic plane must have been undetected in their analyses. The approach proposed by \citet{1998ApJ...506..818D} allowed us to classify the population of the novae in the host galaxy, however, it seems that 
kinematical and/or 
spectroscopic observations for the novae in different Galactic components are needed to clarify the population of 
the Galactic novae with certainty.

In order to understand the nova formation and evolution scenarios, 
key ingredients are Galactic model parameters such as space density and
scale height (and scale length), which also describe the spatial distributions
of these objects in the Galaxy. However, statistical and systematic
uncertainties arising from the small number of systems and uncertain
distances, affect the estimates of Galactic model parameters. In this
study, Galactic model parameters of the Galactic novae were
investigated with fairly homogeneous samples containing a large number of
novae, but it is not free from the selection effect
caused by undiscovered novae. 
However, limiting the volume occupied by novae makes the selection effects less significant. 
According to present samples and their analyses, the scale heights were
calculated to be in the range of  $0.15-0.20$ kpc. 
It is important to conclude that exponential scale height of 0.16~kpc for a mean value is best estimate for the Galactic novae. 
This is consistent with previous estimates, which  depend on a small number of Galactic novae specifically; 
0.150 kpc \citep{1984ApJS...54..443P}
0.125 kpc \citep{1984Ap&SS..99..363D}, 
0.164 kpc \citep{2015NewA...34..234O}. However, these are 
smaller than the scale height of 0.25 kpc estimated by
\citet{2017ApJ...834..196S}.
For cataclysmic variables, the scale height estimates in the range of
$\sim 0.1-0.2$ kpc \citep[]{1984Ap&SS..99..363D, 1984ApJS...54..443P,1996A&A...312...93V,
2008NewA...13..133A, 2008A&A...489.1121R, 2015NewA...34..234O} 
have been made over the years, and our result is in this range
, but it is smaller than the
scale height of the thin disk population which lies in the range  $0.2-0.3$
kpc \citep[e.g.][and references therein]{2005A&A...433..173C,2006NewA...12..234B,2008ApJ...673..864J,2015PASA...32...12Y}. This indicates that the interactions in binary systems, and/or rapid outbursts may accelerate the evolutions of the novae in the Galaxy that cause them to lie closer to the Galactic plane. If so, it raises a question why a considerable number of outbursts at thick disk or halo can not be detected? A possible answer may be that the lifetimes of novae are much smaller than that of the Galactic (thin) disk, or that they change their sub-types of cataclysmic variables from (old) novae to dwarf novae or nova-like systems on relatively short time-scale, or that we just do not detect them due to observational effects.

For the outburst density, we used the Z-distributions of the classical novae within a 
cylinder of a radius $d_{XY}=2$ kpc, and calculated the outburst density as $3.6 \pm 0.25$ and $4.2 \pm 0.26 \times 10^{-10}$ pc$^{-3}$ yr$^{-1}$ for the novae in sample 1 and 2, respectively. The limits of the local outburst density were also calculated as 1.3 and 6 $\times 10^{-10}$ pc$^{-3}$ yr$^{-1}$ by adopting cylinder radii of 1 and 4~kpc, respectively. This result represents very well the outburst density of novae in the solar neighbourhood compared to that obtained in previous studies 
\citep{1984ApJS...54..443P, 1984Ap&SS..99..363D, 1992MNRAS.258..449N, 1993A&A...275..239D, 2017ApJ...834..196S}.
The adopted outburst density allows us to infer a space density of classical novae in the range of 
$\sim1 - 10 \times 10^{-6}$ pc$^{-3}$ when using $T_R=3000-23,000$ yr, while the potential range of the space density was calculated as 0.4~--~16  $\times 10^{-6}$ pc$^{-3}$. 
Since ratios between the recurrence times vary within an order of magnitude, these results are only limits for the actual space density. 
They are consistent with lower limit of  $\sim 10^{-6}-10^{-8}$ pc$^{-3}$ in previous studies 
\citep{1974MNSSA..33...21W, 1984ApJS...54..443P, 1984Ap&SS..99..363D, 1986ApJ...307..170D, 
1993A&A...275..239D}. 
The theoretical population analyses on cataclysmic variables 
predict a similar space  density $10^{-5}-10^{-4}$ pc$^{-3}$ 
\citep{1986A&A...158..161R, 1992A&A...261..188D, 1993A&A...271..149K,1996ApJ...465..338P}.
However, two uncertainties should be taken into consideration in the calculation of the space density. First, the contribution of the RNe systems labeled as CNe effects the mean recurrence time and the space density. 
As we mentioned in the discussion on the MMRD relation, we do not know the actual fraction of RNe in the CNe samples. Even though, we excluded the RNe sample (including RN candidates proposed by \citet{2014ApJ...788..164P}) from our calculations, a possible extra contribution may exists. Besides, we adopted a lower limit of 3000 years for the mean recurrence time  of classical novae, and this does not represent recurrent novae whose recurrence time is in the range of 1 year \citep{2016ApJ...833..149D} up to $\sim 100$ years \citep{2010ApJS..187..275S}. 
If we assume that all novae are actually recurrent novae, the space density can only be calculated by grouping novae with similar recurrence times, but this is not possible with the current Galactic novae sample.
The other but unsolvable problem arises from an observational selection effect.   
The contribution from novae with unrecorded outbursts 
makes the space density results of the novae far from certain. Defining limits and 
using a larger sample in these analyses make the sample approximately complete in a certain volume. Nevertheless,
selection effects on the sample in our study are likely less than previous samples in the literature. 
Thus, Galactic model parameters of the Galactic novae in the solar neighbourhood derived from this sample 
should be more reliable.

Using the local outburst density and scale height obtained in this study, we calculated the nova rate for the Galactic disk in the range of $\sim20$ to $\sim100$ per year using the same methodology as \citet{2017ApJ...834..196S}. The average estimate of the disk nova rate for our samples is $67^{+21}_{-17}$ yr$^{-1}$, which is larger than the best estimate of $45$ yr$^{-1}$ for disk nova rate obtained by \citet{2017ApJ...834..196S}. Recently, \citet{2015ApJS..219...26M} have measured a nova rate $13.8\pm 2.6$ yr$^{-1}$ for the Galactic bulge alone using OGLE observations. By adding this value, the global nova rate can be estimated as $\sim80$ per year. In the literature, the Galactic nova rate is not well estimated and differs in the range between 20 and 260 $yr^{-1}$ 
\citep{1972SvA....16...41S,1994A&A...286..786D, 1997ApJ...487..226S, 2002AIPC..637..462S, 2017ApJ...834..196S}
, but our result is larger than typically adopted estimates for the Galaxy, and also larger than that calculated for the novae in other galaxies \citep{2000ApJ...530..193S, 2006MNRAS.369..257D, 2014ASPC..490...77S}. 
Since nova explosions contribute to the enrichment of the interstellar medium and to the chemical evolution of the Galaxy, the nova rate is essential for Galactic chemical evolution models. 
However, the incompleteness of a nova sample in the Galaxy gives rise to uncertainties on the measurements. The nova rate could be higher when considering the contributions of (undetected/old) novae corresponding to other populations.

\section*{Acknowledgements}

We acknowledge the anonymous referee for helpful comments on this paper. We wish to thank 
Kai Schwenzer for his support on the proof reading.
This work has been  supported in part by the Scientific
and Technological Research Council of Turkey (T\"UB\.ITAK) 114F271. TG was supported in part by the Scientific Research Project Coordination Unit of Istanbul
University (Project no: 53496).  
This research has made use
of the International Variable Star Index (VSX) database, operated at AAVSO,
Cambridge, Massachusetts, USA.
We acknowledge with thanks the variable star observations
from the AAVSO International Database contributed by
observers worldwide which were used in this research.
This research has made use of the SIMBAD, and NASA's Astrophysics Data System Bibliographic Services.

\begin{landscape}
\begin{table}
\centering
\caption{The Galactic novae whose distances are known.
We list Galactic coordinates $l$ and $b$, the amplitude of outburst, the visual maximum magnitude at outburst $V_{max}$, the decline times $t_{2,3}$ as light curve parameters. Type of novae corresponding to their light curve shape \citep[LC; ][]{2010AJ....140...34S}, the spectral type and recurrence type are given. The full width at half maximum (FWHM) is measured from the Balmer lines  at or near outburst maximum. $P_{orb}$ is the orbital period of the nova. The adopted distance ($D$) calculations are given with their measurement methods. $E(B-V)$ gives the adopted reddening estimates obtained from the literature. Numbers in parenthesis are the reference codes.
}
\scriptsize
\begin{tabular}{lcccccccccccccc}
\hline
\multicolumn{15}{c}{1a. The novae that are used in MMRD analyses.}\\
\hline
\multirow{2}{*}{Nova}	&$l$	&$b$	&Amplitude$^{a}$	&$V_{max}$	&$t_{2}$	&$t_{3}$	&\multirow{2}{*}{LC type} &\multirow{2}{*}{Spec type} &\multirow{2}{*}{RN type} &FWHM &$P_{orb}$	&$D$	&\multirow{2}{*}{Method$^{b}$} &$E(B-V)$\\
 &(deg)	&(deg) & (mag) &(mag)	&(days) 	&(days)	&	&	& &(km/s)	 &(days)	&(kpc)	&	&(mag)\\
\hline
CI Aql  & 31.688 & -0.812 & 7.7 (V) (1) & 9 (1) & 25 (1) & 32 (1) & P (1) &   & RN (2) & 2250 (3) & 0.618 (4) & $3.9\pm0.5$ & rdr  (5) & $0.92\pm0.22$ (6)\\ 
V500 Aql  & 47.608 & -9.462 & 11.2 (p) (4) & 6.1 (7) & 17 (7) & 42 (7) &   & He/N? (7) &   & 1400 (8) & 0.145 (4) & $5.9\pm1.2$ & epar  (9) & $0.16\pm0.03$ (10)\\ 
V603 Aql  & 33.164 & 0.829 & 12.2 (V) (1) & -0.5 (1) & 5 (1) & 12 (1) & O (1) & hybrid (1) &   &   & 0.138 (11) & $0.34\pm0.06$ & tpar  (12) & $0.08\pm0.02$ (13)\\ 
V1229 Aql  & 40.537 & -5.437 & 11.5 (V) (1) & 6.6 (1) & 18 (1) & 32 (1) & P (1) & Fe II (7) & cRN: (14) &   &   & $2.4\pm0.6$ & epar  (9) & $0.5\pm0.08$ (15)\\ 
V1494 Aql  & 40.974 & -4.742 & 13 (V) (1) & 4.1 (1) & 8 (1) & 16 (1) & O (1) & Fe II (16) &   & 2470 (17) & 0.135 (4) & $1.2\pm0.2$ & epar  (18) & $0.6\pm0.1$ (19)\\
\hline
\end{tabular}
\label{table1}
\end{table}
\end{landscape}

\begin{landscape}
\begin{table}
\contcaption{}
\scriptsize
\centering
\begin{tabular}{lcccccccccccccc}
\hline
\multirow{2}{*}{Nova}	&$l$	&$b$	&Amplitude$^{a}$	&$V_{max}$	&$t_{2}$	&$t_{3}$	&\multirow{2}{*}{LC type} &\multirow{2}{*}{Spec type} &\multirow{2}{*}{RN type} &FWHM &$P_{orb}$	&$D$	&\multirow{2}{*}{method$^{b}$} &$E(B-V)$\\
 &(deg)	&(deg) & (mag) &(mag)	&(days) 	&(days)	&	&	& &(km/s)	 &(days)	&(kpc)	&	&(mag)\\
\hline
V1324 Sco  & 357.426 & -2.872 & <10  (V) (37) & 10.14 (100) & 25 (100) & 29.5 (100) &   & Fe II (74) &   & 2000 (74) & 0.133 (101) & :$4.3\pm0.9$ & rdr  (6) & $1.16\pm0.12$ (102)\\ 
V496 Sct  & 25.284 & -1.768 & <11.9  (V) (37) & 7.07 (103) & 59 (103) & 90 (103) &   & Fe II (103) &   & 1200 (103) &   & $3.2\pm0.8$ & rdr  (6) & $0.57\pm0.06$ (6)\\ 
FH Ser  & 32.909 & 5.786 & 12.3 (V) (1) & 4.5 (1) & 49 (1) & 62 (1) & D (1) & Fe II (7) &   &   &   & $0.95\pm0.05$ & epar  (61) & $0.6\pm0.1$ (6)\\ 
LW Ser  & 12.959 & 6.047 & 11.1 (V) (1) & 8.3 (1) & 32 (1) & 52 (1) & D (1) &   &   &   &   & $2.34\pm0.46$ & epar  (62) & $0.39\pm0.1$ (6)\\ 
V732 Sgr$^{d}$ & 2.53 & -1.188 & $\sim$9.6  (V) (1) & 6.4 (1) & 65 (1) & 75 (1) & D (1) &   &   &   &   & $3.0\pm0.2$ & rdr  (6) & $0.81\pm0.16$ (104)\\ 
V3890 Sgr  & 9.204 & -6.443 & 7.4 (V) (1) & 8.1 (1) & 6 (1) & 14 (1) & S (1) & hybrid (21) & RN (2) & 2140 (105) & 519.7 (92) & $7.0\pm1.6$ & comp.  (2) & $0.9\pm0.3$ (2)\\
\hline
\end{tabular}
\end{table}
\end{landscape}

\begin{landscape}
\begin{table}
\contcaption{}
\scriptsize
\centering
\begin{tabular}{lcccccccccccccc}
\hline
\multirow{2}{*}{Nova}	&$l$	&$b$	&Amplitude$^{a}$	&$V_{max}$	&$t_{2}$	&$t_{3}$	&\multirow{2}{*}{LC type} &\multirow{2}{*}{Spec type} &\multirow{2}{*}{RN type} &FWHM &$P_{orb}$	&$D$	&\multirow{2}{*}{method$^{b}$} &$E(B-V)$\\
 &(deg)	&(deg) & (mag) &(mag)	&(days) 	&(days)	&	&	& &(km/s)	 &(days)	&(kpc)	&	&(mag)\\
\hline
V4444 Sgr$^{c}$ & 3.818 & -3.389 & >13.4  (V) (1) & 7.6 (1) & 5 (1) & 9 (1) & S (1) &   & cRN? (14) & 800 (155) &   & $2.27\pm0.58$ & rdr  (6) & $0.45\pm0.1$ (6)\\ 
V4633 Sgr$^{c}$ & 5.128 & -6.231 & 11.3 (V) (1) & 7.4 (1) & 17 (1) & 44 (1) & P (1) & Fe II (114) & cRN: (14) & 1700 (114) & 0.12557 (156) & :$1.36\pm0.46$ & rdr  (6) & $0.26\pm0.05$ (6)\\ 
V5113 Sgr$^{c, d, e}$ & 3.721 & -4.091 & <9.2  (V) (37) &   &   &   &   &   &   & 550 (157) &   & :$0.95\pm0.21$ & rdr  (6) & $0.1\pm0.02$ (158)\\ 
V5115 Sgr$^{c, d}$ & 6.046 & -4.567 & >10.1  (V) (1) & 7.9 (1) & 7 (1) & 13 (1) & S (1) & He/N (21) &   & 1300 (159) &   & $\sim$$3.0\pm1.0$ & rdr  (6) & $0.53\pm0.05$ (160)\\ 
V5116 Sgr$^{c}$ & 2.136 & -6.833 & >8.4  (V) (1) & 7.6 (1) & 12 (1) & 26 (1) & S (1) & Fe II (21) &   & 970 (161) & 0.1238 (162) & :$1.55\pm0.7$ & rdr  (6) & $0.23\pm0.06$ (6)\\ 
V5117 Sgr$^{d, e}$ & 354.624 & -6.377 & <7.8  (V) (37) & 8.6 (163) & 59 (164) & 83 (165) &   & Fe II (74) &   & 1600 (74) &   & $1.45\pm0.33$ & rdr  (6) & $0.5\pm0.1$ (166)\\
\end{tabular}
\end{table}
\end{landscape}

\begin{landscape}
\begin{table}
\caption{The Galactic novae whose distances could not be estimated. Lower limits of distances for the novae, calculated from reddening-distance relations, are given as well. The column definitions are the same as in Table \ref{table1}.}
\scriptsize
\centering
\begin{tabular}{lccccccccccccc}
\hline
\multicolumn{14}{c}{ 2a. The novae with well-determined light curve parameters.}\\
\hline
\multirow{2}{*}{Nova}	&$l$	&$b$	&Amplitude$^{a}$	&$V_{max}$	&$t_{2}$	&$t_{3}$	&\multirow{2}{*}{LC type} &\multirow{2}{*}{Spec type} &\multirow{2}{*}{RN type} &FWHM &$P_{orb}$	&$D$	&$E(B-V)$\\
 &(deg)	&(deg) & (mag) &(mag)	&(days) 	&(days)	&	&	& &(km/s)	 &(days)	&(kpc)	&(mag)\\
\hline
OS And & 106.051 & -12.117 & 11 (V) (1) & 6.5 (1) & 11 (1) & 23 (1) & D (1) &   &   & 1740 (2) &   &    & $0.15\pm0.05$ (3)\\ 
DO Aql & 31.705 & -11.806 & 9.5 (V) (1) & 8.5 (1) & 295 (1) & 900 (1) & F (1) &   &   &   & 0.16776 (4) &    &  \\ 
V1370 Aql & 38.813 & -5.947 & 10.3 (V) (1) & 7.7 (1) & 15 (1) & 28 (1) & D (1) &   &   & 730 (5) &   & $\geq1.6$  (6) & $0.35\pm0.05$ (7)\\ 
V1419 Aql & 36.811 & -4.1 & $\sim$13.4 (V) (1) & 7.6 (1) & 25 (1) & 32 (1) & D (1) &   &   &   &   & $\geq4.7$  (6) & $0.5\pm0.05$ (3)\\ 
V1425 Aql & 33.011 & -3.893 & $\sim$8 (V) (1) & 8.0 (1) & 27 (1) & 79 (1) & S (1) & Fe II (1) &   & 1340 (8) & 0.2558 (4) & $\geq2.2$  (6) & $1.0\pm0.3$ (9)\\
\hline
\end{tabular}
\label{table2}
\end{table}
\end{landscape}

\begin{landscape}
\begin{table}
\contcaption{}
\scriptsize
\centering
\begin{tabular}{lccccccccccccc}
\hline
\multirow{2}{*}{Nova}	&$l$	&$b$	&Amplitude$^{a}$	&$V_{max}$	&$t_{2}$	&$t_{3}$	&\multirow{2}{*}{LC type} &\multirow{2}{*}{Spec type} &\multirow{2}{*}{RN type} &FWHM &$P_{orb}$	&$D$	&$E(B-V)$\\
 &(deg)	&(deg) & (mag) &(mag)	&(days) 	&(days)	&	&	& &(km/s)	 &(days)	&(kpc)	&(mag)\\
\hline
V4743 Sgr & 14.124 & -11.877 & 11.8 (V) (1) & 5.0 (1) & 6 (1) & 12 (1) & S (1) & Fe II (13) & cRN: (10) & 2400 (69) & 0.281 (70) &    & $0.25$ (71)\\ 
V4745 Sgr & 1.493 & -12.382 & >9.7 (V) (1) & 7.3 (1) & 79 (1) & 190 (1) & J (1) & Fe II (13) &   & 1600 (72) & 0.20782 (73) &    & $0.1$ (74)\\ 
V5114 Sgr & 3.943 & -6.312 & >12.9 (V) (1) & 8.1 (1) & 9 (1) & 21 (1) & S (1) & Fe II (35) &   & 2000 (75) &   & $\geq0.99$  (45) & $0.6\pm0.3$ (75)\\ 
V5583 Sgr & 358.1 & -6.394 & <12.7 (V) (4) & 7.43 (76) & 4.97 (77) & 9.52 (77) &   & He/N (77) &   & 2300 (76) &   &    &  \\ 
V5855 Sgr & 3.981 & -4.025 & <14.3 (V) (4) & 7.54 (46) & 18 (46) & 23.5 (46) &   & Fe II (46) &   &   &   & $\geq2.3$  (45) & $0.5$ (46)\\ 
V5856 Sgr & 4.293 & -6.461 &   & 5.9 (46) & 11.3 (46) & 14.5 (46) &   & Fe II (78) &   &   &   &    & $1.02$ (46)\\
\hline
\end{tabular}
\end{table}
\end{landscape}

\begin{landscape}
\begin{table}
\contcaption{}
\scriptsize
\centering
\begin{tabular}{lccccccccccccc}
\hline
\multirow{2}{*}{Nova}	&$l$	&$b$	&Amplitude$^{a}$	&$V_{max}$	&$t_{2}$	&$t_{3}$	&\multirow{2}{*}{LC type} &\multirow{2}{*}{Spec type} &\multirow{2}{*}{RN type} &FWHM &$P_{orb}$	&$D$	&$E(B-V)$\\
 &(deg)	&(deg) & (mag) &(mag)	&(days) 	&(days)	&	&	& &(km/s)	 &(days)	&(kpc)	&(mag)\\
\hline
DY Pup & 245.823 & 4.362 & 12.6 (V,p) (14) & 7.0 (15) & 118 (84) & 160 (15) &   &   &   &   & 0.139 (4) &    &  \\ 
HS Pup & 247.756 & -2.112 & 10.8 (B) (14) & 8.0 (15) & 33 (84) & 65 (15) &   &   &   &   & 0.2671 (89) &    &  \\ 
HZ Pup & 246.179 & 1.385 & 9.3 (B) (14) & 7.7 (15) & 35 (84) & 70 (15) &   &   &   &   & 0.213 (4) &    &  \\ 
V597 Pup & 252.53 & 0.545 & 13.8 (V) (4) & 7.0 (121) & 2.5 (122) &   &   & He/N (122) &   & 1800 (123) & 0.11119 (124) & $\geq8.0$  (6) & $0.3$ (125)\\ 
V598 Pup & 249.108 & -13.814 & 13 (V) (4) & 3.46 (29) & 9 (126) &   &   &   &   &   &   &    & $0.09\pm0.08$ (126)\\ 
KP Sco & 354.058 & -3.297 & 11.6 (j,p) (14) & 9.4 (15) & 21 (84) & 38 (15) &   &   &   &   &   &    &  \\
\hline
\end{tabular}
\end{table}
\end{landscape}

\begin{landscape}
\begin{table}
\contcaption{}
\scriptsize
\centering
\begin{tabular}{lccccccccccccc}
\hline
\multirow{2}{*}{Nova}	&$l$	&$b$	&Amplitude$^{a}$	&$V_{max}$	&$t_{2}$	&$t_{3}$	&\multirow{2}{*}{LC type} &\multirow{2}{*}{Spec type} &\multirow{2}{*}{RN type} &FWHM &$P_{orb}$	&$D$	&$E(B-V)$\\
 &(deg)	&(deg) & (mag) &(mag)	&(days) 	&(days)	&	&	& &(km/s)	 &(days)	&(kpc)	&(mag)\\
\hline
V5668 Sgr & 5.38 & -9.867 & 11.9 (V) (4) & 4.32 (148) & 100 (148) &   &   & Fe II (148) &   &   &   &    &  \\ 
V630 Sgr & 357.768 & -6.967 &   & 1.6 (15) & 4 (15) & 11 (15) &   &   &   &   &   &    &  \\ 
V726 Sgr & 5.479 & -5.519 &   & 10.5 (51) & 45 (84) & 95 (51) &   &   &   &   & 0.41141 (149) &    &  \\ 
V737 Sgr & 2.531 & -3.982 & 8.7 (j,p) (14) & 10.7 (51) &   & >70  (51) &   &   &   &   &   &    &  \\ 
V787 Sgr & 0.235 & -3.495 & 11.2 (j,p) (14) & 9.8 (15) & 23 (84) & 74 (15) &   &   &   &   &   &    &  \\ 
V909 Sgr & 358.767 & -10.403 & 13.2 (j,p) (14) & 6.8 (15) & 3.8 (84) & 7 (15) &   &   &   &   & 0.14286 (89) &    &  \\
\end{tabular}
\end{table}
\end{landscape}

\bsp 
\label{lastpage} 
\end{document}